\documentclass[prx,floatfix,amsmath,twocolumn,notitlepage,superscriptaddress,nofootinbib]{revtex4-1}

\usepackage{amssymb}
\usepackage{graphicx}
\usepackage{graphics}
\usepackage{amsmath}
\usepackage{amsthm}
\usepackage{color}
\usepackage{dsfont}

\usepackage{float}
\usepackage{mathtools}
\usepackage[colorlinks=true, linkcolor=blue, citecolor=red]{hyperref}
\usepackage{verbatim}
\usepackage{xcolor}

\newcommand{\calc}{\mathcal{C}}

\newcommand{\e}{\mathcal{E}}
\usepackage{color}

\newcommand{\super}{{\rm SC}}

\newcommand{\p}{\mathbf{p}}
\newcommand{\q}{\mathbf{q}}

\def\>{\rangle}
\def\<{\langle}

\def\id{\mathsf{id}}

\def\mE{\mathcal{E}}

\def\mF{\mathcal{F}}
\def\mN{\mathcal{N}}

\def\mL{\mathcal{L}}

\renewcommand{\qedsymbol}{\nobreak \ifvmode \relax \else
	\ifdim \lastskip<1.5em \hskip-\lastskip \hskip1.5em plus0em
	minus0.5em \fi \nobreak \vrule height0.75em width0.5em
	depth0.25em\fi}

\renewcommand{\geq}{\geqslant}
\renewcommand{\leq}{\leqslant}
\newcommand{\post}{{\rm post}}
\newcommand{\pree}{{\rm pre}}

\def\mb{\mathfrak{B}}

\def\md{\mathfrak{D}}

\def\mq{\mathfrak{Q}}

\def\x{\mathbf{x}}

\newcommand{\locc}{{\rm LOCC}}

\newcommand{\losr}{{\rm LOSR}}
\newcommand{\povm}{{\rm POVM}}

\def\ml{\mathfrak{L}}

\newcommand{\bl}{{\rm BL}}

\newtheorem{theorem}{Theorem}
\newtheorem*{theorem*}{Theorem}

\newtheorem*{lemma*}{Lemma}

\newtheorem{definition}{Definition}[section]
\newtheorem*{definition*}{Definition}

\theoremstyle{definition}

\theoremstyle{remark}
\newtheorem*{remark}{Remark}

\theoremstyle{definition}

\newcommand{\bea}{\begin{eqnarray}}
\newcommand{\eea}{\end{eqnarray}}
\newcommand{\be}{\begin{equation}}
\newcommand{\ee}{\end{equation}}
\newcommand{\ba}{\begin{equation}\begin{aligned}}
\newcommand{\ea}{\end{aligned}\end{equation}}

\def\be{\begin{equation}}
\def\ee{\end{equation}}

\newcommand{\cptp}{{\rm CPTP}}

\newcommand{\lose}{{\rm LOSE}}

\newcommand{\mM}{\mathcal{M}}

\newcommand{\lr}{\rangle\langle}
\newcommand{\la}{\langle}
\newcommand{\ra}{\rangle}
\newcommand{\tr}{{\rm Tr}}
\newcommand{\pr}{\mathrm{p}}

\newcommand{\ve}[1]{{\left\vert\kern-0.25ex\left\vert\kern-0.25ex\left\vert #1 
    \right\vert\kern-0.25ex\right\vert\kern-0.25ex\right\vert}}

\newcommand{\mbf}[1]{\mathbf{#1}}
\newcommand{\mbb}[1]{\mathbb{#1}}

\newcommand{\ket}[1]{|#1\rangle}

\newcommand{\eqdef}{\coloneqq}

\def\tA{\tilde{A}}
\def\tB{\tilde{B}}

\def\herm{{\rm Herm}}

\def\chsh{{\rm CHSH}}

\definecolor{cool_green}{rgb}{0.0, 0.5, 0.0}

\usepackage{tikz,lipsum,lmodern}
\usepackage[most]{tcolorbox}
  
\begin{document}
	

	\title{Quantum Bell Nonlocality is Entanglement}
	
\author{Kuntal Sengupta}\email{kuntal.sengupta.in@gmail.com}
\affiliation{Department of Mathematics and Statistics, Institute for Quantum Science and Technology,
University of Calgary, AB, Canada T2N 1N4}

\author{Rana Zibakhsh}\email{rana.zibakhshshabgah@ucalgary.ca}
\affiliation{Department of Physics and Astronomy, University of Calgary, AB, Canada T2N 1N4} 

\author{Eric Chitambar}\email{echitamb@illinois.edu}
\affiliation{Department of Electrical and Computer Engineering, Coordinated Science Laboratory, University of Illinois at Urbana-Champaign,  Urbana,  IL 61801}

\author{Gilad Gour}\email{giladgour@gmail.ca}
\affiliation{Department of Mathematics and Statistics, Institute for Quantum Science and Technology,
University of Calgary, AB, Canada T2N 1N4}
\affiliation{Department of Physics and Astronomy, University of Calgary, AB, Canada T2N 1N4}

	\date{\today}
	
	\begin{abstract}
Bell nonlocality describes a manifestation of quantum mechanics that cannot be explained by any local hidden variable model.  Its origin lies in the nature of quantum entanglement, although understanding the precise relationship between nonlocality and entanglement has been a notorious open problem.  In this paper, we resolve this problem by developing a dynamical framework in which quantum Bell nonlocality emerges as special form of entanglement, and both are unified as resources under local operations and classical communication (LOCC).  Our framework is built on the notion of quantum processes, which are abstract quantum channels mapping elements between fixed intervals in space and time.  Entanglement is then identified as a quantum process that cannot be generated by LOCC while Bell nonlocality is the subset of these processes that have an \textit{instantaneous} input-output delay time. LOCC pre-processing is a natural set of free operations in this theory, thereby enabling all entangled states to activate some form of Bell nonlocality. In addition, we generalize the CHSH witnesses from the state domain to the domain of entangled quantum measurements, and provide a systematic method to quantify the Bell nonlocality of a bipartite quantum channel.
\end{abstract}

	\maketitle

\section{Introduction}
 
Entanglement and Bell nonlocality represent two of the most stunning non-classical features of quantum mechanics.  Entanglement is traditionally understood as the property of non-separability \cite{Werner-1989a, Horodecki-2009a}.  More precisely, a bipartite quantum state $\rho^{AB}$ is entangled if it cannot be separated into a statistical mixture of product states,
\begin{equation}
\label{Eq:entangled}
\rho^{AB}\not=\sum_\lambda p(\lambda)\rho_\lambda^A\otimes\rho_\lambda^B.
\end{equation}
In contrast, quantum Bell nonlocality is the property that enables certain quantum states to generate nonlocal classical channels.  The latter refers to classical channels that cannot be described by a local hidden variable (LHV) model; i.e the transition probabilities $W(a,b|x,y)$ cannot be separated into a statistical mixture of product channels, 
\begin{equation}
\label{Eq:Bell-nonlocal}
W(a,b|x,y)\not=\sum_\lambda p(\lambda) W_\lambda(a|x)W_\lambda(b|y).
\end{equation}
Building on the pioneering insight of J.S. Bell \cite{Bell-1964a}, one can test whether a given channel satisfies a LHV model by checking whether the probabilities $W(a,b|x,y)$ satisfy a finite family of inequalities, known as Bell inequalities \cite{Brunner-2014a}.  The well-known Clauser-Horne-Shimony-Holt (CHSH) Inequality is one such inequality \cite{Clauser-1969a}, and it completely characterizes the set of nonlocal channels for binary inputs and outputs \cite{Froissart-1981a, Fine-1982a}.  

The most direct way to generate a nonlocal channel from a quantum state $\rho^{AB}$ is by performing local measurements. The channel generated by such a process has transition probabilities determined by Born's rule,
\begin{equation}
\label{Eq:Born}
W(a,b|x,y)=\tr[\rho^{AB}(M^{x}_a\otimes N^y_b)],
\end{equation}
where $\{M^x_a\}_a$ represent measurement operators on Alice's system ($A$) for measurement choice $x$ and outcome $a$, and similarly for the operators $\{N^y_b\}_b$ on Bob's system ($B$).  For example, by choosing suitable pairs of local measurements, a bipartite maximally entangled state $\ket{\phi^+}=\sqrt{1/2}(\ket{00}+\ket{11})$ can be used to  generate a classical channel that violates the CHSH Inequality.   

The similarity between Eqns. \eqref{Eq:entangled} and \eqref{Eq:Bell-nonlocal} suggests that quantum entanglement and Bell nonlocality can perhaps be unified in some fundamental way.  By inspection, we can conclude that entangled states and nonlocal classical channels are both objects that cannot be built using local operations and shared randomness (LOSR).  This has led some to conclude that LOSR should provide the underlying fabric to weave together entanglement and Bell nonlocality \cite{Schmid-2020b}.  However, as we argue below and throughout this paper, a restriction to LOSR misses key features of both entanglement and nonlocality.   Another approach is needed.

\begin{figure}[t]
\centering
\includegraphics[width=.48\textwidth]{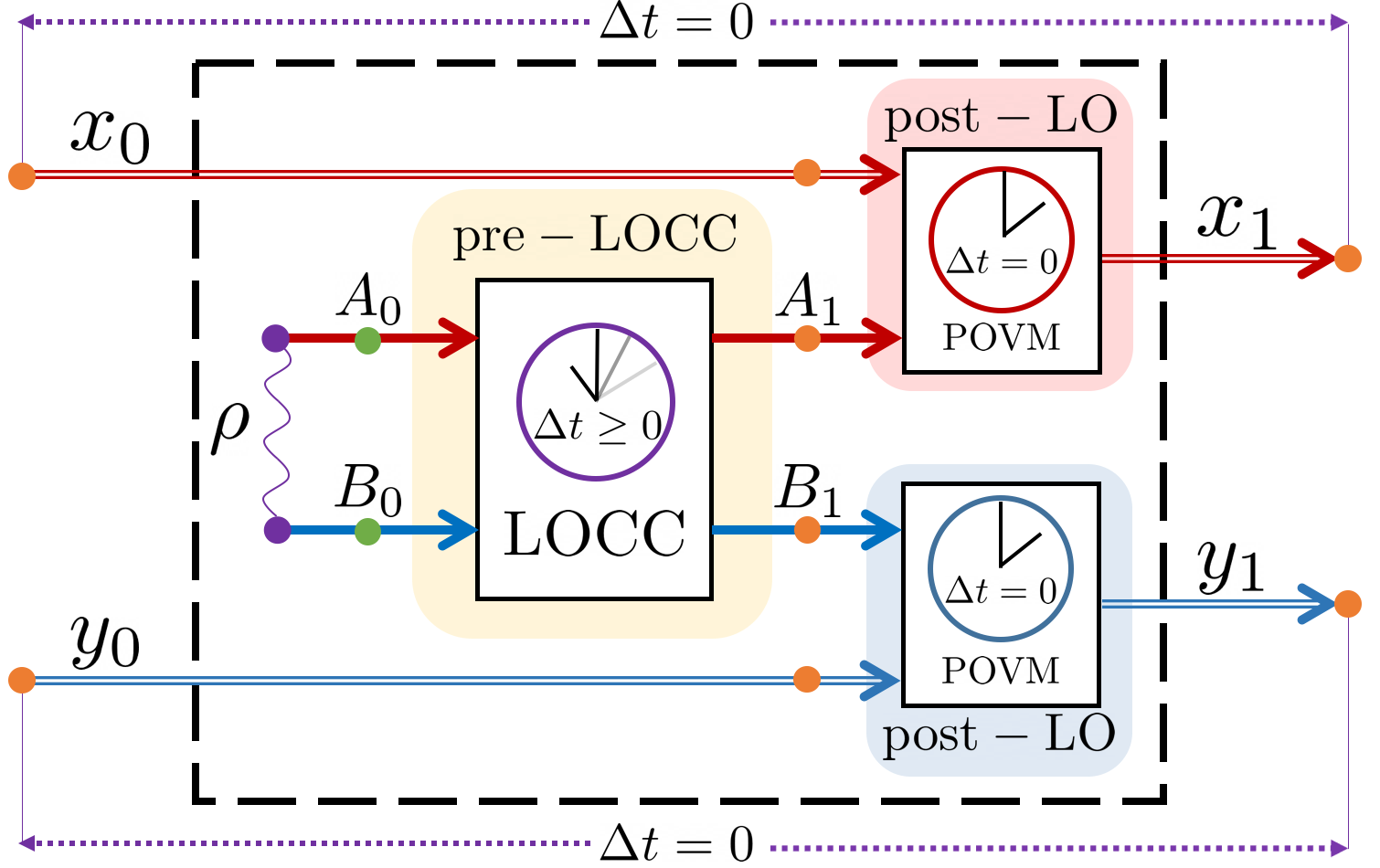}
 \caption{\linespread{1}\selectfont{\small A state $\rho$ that cannot violate a Bell inequality can be transformed into a state that violates one by performing a pre-LOCC map.  The key property we identify in this work is that \textit{after} the LOCC map, what remains is a bipartite classical channel with essentially zero delay time between the inputs $(x_0,y_0)$ and outputs $(x_1,y_1)$. Dots with the same colour correspond to the same time.}}
 \label{bell_scenario}
\end{figure}

While bipartite entanglement is well understood~\cite{Horodecki-2009a}, its relationship to Bell nonlocality is far more elusive ~\cite{Werner-1989a, Hardy-1993a, Eberhard-1993a, Popescu-1995a, Barrett-2002a, Wiseman-2007a, Methot-2007a, Masanes-2008a, Hirsch-2013a, Quintino-2015a, Augusiak-2015a, Hirsch-2016a}.  Most notable is the existence of entangled states that cannot violate \textit{any} Bell inequality when generating a classical channel according to Eq. \eqref{Eq:Born} \cite{Werner-1989a, Barrett-2002a}.  The discovery of such states seems to cast some doubt on whether it is possible to formulate any unified theory of both entanglement and Bell nonlocality.  At the same time, however, quantum Bell nonlocality is known to demonstrate the remarkable effect of super-activation.  There exists two states $\rho^{A_0B_0}$ and $\sigma^{A_0'B_0'}$, each of which is unable to violate a Bell inequality.  Yet when combined and measured jointly, the state $\rho^{A_0B_0}\otimes\sigma^{A_0'B_0'}$ can violate some Bell inequality \cite{Palazuelos-2012a}.  An even more important collection of results for the purposes of this work shows that certain states have hidden nonlocality that becomes revealed through the use of local filters \cite{Popescu-1995a, Hirsch-2013a}.  In fact, \textit{every} bipartite entangled state $\rho^{A_0B_0}$ can violate the CHSH Inequality when combined with another state $\sigma^{A_0'B_0'}$ that itself cannot violate the CHSH Inequality, after the two are allowed to undergo processing by local operations and classical communication (LOCC) prior to receiving any classical inputs for the resultant channel (see Fig. \ref{bell_scenario}) \cite{Masanes-2008a}.  That is, a CHSH-violating classical channel can be generated by $\rho^{A_0B_0}\otimes\sigma^{A_0'B_0'}$ when Eq. \eqref{Eq:Born} is modified to have the form
\begin{align}
\label{Eq:pre-LOCC}
&W(x_1,y_1|x_0,y_0)\nonumber\\
&=\tr\left[\mathcal{L}_{\text{pre}}(\rho^{A_0B_0}\otimes\sigma^{A_0'B_0'})(M_{x_1}^{x_0}\otimes N^{y_0}_{y_1})\right],
\end{align}
where $\mathcal{L}_{\text{pre}}$ is a so-called pre-processing LOCC map that is applied to $\rho^{A_0B_0}\otimes\sigma^{A_0'B_0'}$ before the choice of measurements $(x_0,y_0)$.  A similar result also holds for multipartite entangled states \cite{Liang-2012a}.

LOCC is the natural class of operations to consider when studying quantum entanglement since it models quantum information processing in the ``distant lab'' setting \cite{Plenio-2007a}.  The fact that LOCC is needed to fully reveal the nonlocal features of quantum states, as in Eq. \eqref{Eq:pre-LOCC}, indicates that LOCC is somehow also fundamental to quantum Bell nonlocality.  However, a precise and physically-motivated connection between the theories of quantum entanglement and Bell nonlocality has thus far been lacking.  Here we address this void by describing a framework in which Bell nonlocality emerges as a subtheory of entanglement and both ultimately belong to the same species, resources under LOCC.

We establish our claim within the domain of dynamical resource theories (DRTs), which have been extensively studied in recent years~\cite{PhysRevLett.123.150401,PhysRevResearch.1.033169,Wang_2019,gour2019entanglement,buml2019resource,PhysRevLett.124.100501,PhysRevResearch.2.012035,liu2019resource,gour2018entropy,Kaur_2017,katariya2020geometric,fang2019geometric}.   In fact, the whole field of quantum Shannon theory~\cite{wilde_2013} can be seen as a DRT~\cite{Devetak_2008}.  In a general quantum resource theory, one identifies a restricted subset of quantum operations as being ``free'', and objects that cannot be generated by these free operations are deemed to possess a resource \cite{Coecke-2016a, Chitambar-2019a}.  The essential difference between a static and a dynamic resource theory is the element of time.  While a quantum state represents a quantum system at a given snapshot in time, a quantum channel typically requires a non-zero time for any signal to propagate from the inputs to the outputs.  This delay time is particularly relevant for bipartite channels shared between two spatially-separated parties since the finite speed of information propagation limits how fast the input information of one party can influence the output information of the other.  

The input-output delay time of bipartite channels has been a notorious consideration in almost all Bell experiments.  The so-called locality loophole describes how the CHSH Inequality can be violated under a local hidden variable model when the measurement settings on Alice's side can propagate to Bob's laboratory in a time shorter than the input-output delay time of Bob's measurement device \cite{Aspect-1982a, Bell-2004a}.  For this reason, when modelling an ideal Bell experiment it is usually assumed that Alice's measurement choice (i.e. her channel input $x_0$) is spacelike separated from Bob's measurement outcome (i.e. his channel output $y_1$), and vice-versa.  Hence, what is crucial here is not that Alice and Bob are able to generate the particular bipartite classical channel capable of violating the CHSH Inequality, but rather that they are able to generate this channel using local physical processes having input-output delay times shorter than the time it takes light to travel between the two laboratories.  While previous efforts to construct a resource theory of Bell nonlocality \cite{Gallego-2012a, deVicente-2014a, Horodecki-2015a, Gallego-2017a, Wolfe-2020a, Schmid-2020a} may have implicitly acknowledged this crucial fact or used it to motivate the set of free operations, we take the approach of building the input-output delay time directly into the resource theory.

The framework introduced here considers every quantum channel on the abstract level as a completely-positive trace-preserving (CPTP) map between an input and output system separated by a specific fixed distance and time interval.  Hence every quantum channel is associated with a family of dynamical objects, which we refer to as quantum processes and which differ in their input-output spacetime separations.  Quantum processes constitute the resource objects in this theory, and the free operations are LOCC superprocesses, which are LOCC-implementable protocols that map one process to another.  Accounting for the input-output delay time of a channel allows us to draw a distinction between two dynamical resources under LOCC: \textit{ instantaneous} and \textit{non-instantaneous}.  Using this distinction, one can identify quantum Bell nonlocality as an \textit{instantaneous classical process} that can be obtained by LOCC only when Alice and Bob are initially supplied with an entangled state, as in Fig. \ref{bell_scenario}.  One of the key features of our model is that pre-LOCC maps are naturally free operations; i.e. static-to-dynamic conversions having the form of Eq. \eqref{Eq:pre-LOCC} are allowed in this framework.  At the same time, the use of LOCC maps \textit{after} the classical inputs $(x_0,y_0)$ in Eq. \eqref{Eq:pre-LOCC} is automatically prohibited as it leads to a non-instantaneous resource.  

In this paper we develop the theory of quantum processes and study interconversions among them.  When extended to bipartite processes and LOCC interconversions, what emerges is a resource theory that encompasses both entanglement and Bell nonlocality.   Within this resource theory, we show that all entangled states and all entangled bipartite channels can exhibit some features of Bell nonlocality using LOCC.  We also introduce a systematic method to quantify the Bell nonlocality of bipartite quantum states and channels within the LOCC paradigm. In addition, in the appendix we generalize the CHSH witnesses from the state domain to the domain of entangled quantum measurements.

\subsection{The Shortcomings of an LOSR Theory}  

Before we unpack our framework in more detail, let us comment on the role that LOSR plays in our theory and explain why LOSR itself is insufficient to formulate a physically motivated resource theory of Bell nonlocality, despite claims to the contrary \cite{Schmid-2020b}.  Unlike LOCC, LOSR lacks a clear operational interpretation when understood in the context of quantum information protocols.  LOSR is typically justified using a common-cause model~\cite{Wolfe-2020a} in which a helper distributes shared randomness to spatially separated parties.  This common randomness, however, is useless for carrying out some protocol without the parties first agreeing on some pre-established strategy, an agreement which inevitably will require some communication.  For example, the only way that Alice, Bob, and perhaps some other party, can recognize which physical system encodes the shared randomness is by prior interactive communication.  A second issue is that LOSR misses certain important aspects of Bell nonlocality such as it being ``hidden'' in some states; i.e. certain channels generated in Eq. \eqref{Eq:pre-LOCC} cannot also be realized when the pre-LOCC map $\mathcal{L}_{\text{pre}}$ is restricted to LOSR.  The framework presented here overcomes these problems since it starts with the physically-motivated class of LOCC and then arrives at a restricted subset of operations, which includes pre-LOCC, by incorporating the practically-relevant property of input-output delay time.  LOSR still plays an important role in this theory; however it emerges not by appealing to some common cause, but rather by considering bipartite LOCC channels that are implemented with zero input-output delay time.

\subsection{Notations}

In this paper, quantum physical systems and their corresponding Hilbert spaces will be denoted by $A$ and $B$, while classical systems will be denoted by $X$ and $Y$. We will make a distinction between \emph{static} vs \emph{dynamic} systems.
Static systems will be denoted with subscripts such as $A_0,\;A_1,\;B_0,\;B_1$. 
Dynamic systems will be denoted without subscripts; for example, $A$ refers to an input-output system $(A_0, A_1)$, where the zero subscript will always refer to the input subsystem and the subscript one to the output subsystem. Similarly, $B=(B_0,B_1)$, $X=(X_0,X_1)$, etc. One exception of this notation is the auxiliary systems (e.g. environment system, reference systems) $E$ and $R$ which will always correspond to static systems.

 We will also use the notation $A_1B_1 \equiv A_1 \otimes B_1$ to represent a bipartite (static) system. We use the notation $|A|=|A_0||A_1|$ to denote the dimension of a dynamic system $A$, where $|A_0|$ and $|A_1|$ are the dimensions of its corresponding input and output subsystems. The space of bounded linear operators describing endomorphisms on  $A_1$ is denoted by $\mb (A_1)$. We denote by $\mathfrak{D}(A_1)\subset\mb(A_1)$ the set of all density matrices (i.e. positive semidefinite matrices with trace one) acting on $A_1$. 
 As customary, we will use $\rho$ and $\sigma$ to represent mixed quantum states and $\psi$ and $\phi$ to represent pure states. A maximally entangled state in $\md(A_1B_1)$ is denoted by $\phi_+^{A_1B_1}$. 
 
We denote by $\ml(A)\eqdef\ml(A_0\to A_1)$ the set of all linear maps from $\mathfrak{B}(A_0)$ to $\mathfrak{B}(A_1)$. The set of all completely positive and trace preserving (CPTP) maps in $\ml(A)$ is denoted by $\cptp(A)\eqdef\cptp(A_0\to A_1)$. Similarly, we will use the notation $\cptp(AB)$ in short for $\cptp(A_0B_0\to A_1B_1)$. Quantum channels will be denoted with calligraphic letters $\mM,\mN,\mE,\mF$. We will use the superscripts $\mN^A$ and $\mN^{AB}$ to indicate $\mN\in\cptp(A)$ and $\mN\in\cptp(AB)$, respectively. The identity channel in $\cptp(A_0\to A_0)$ is denoted by $\id^{A_0}$. 

Linear maps from $\ml(A)$ to $\ml(B)$ are called supermaps~\cite{Gour-2019a}. Supermaps that takes quantum channels to quantum channels in a complete sense (i.e. even when tensored with the identity supermap) are called superchannels \cite{Chiribella-2008a, Gour-2019a}. The collection of all superchannels from $\ml(A)$ to $\ml(B)$ will be denoted by $\super(A \to B)$. We use the letter $\Theta, \Upsilon$ to denote superchannels. In \cite{Chiribella-2008a,Gour-2019a} it was shown  
that the action of every superchannel $\Theta\in\super(A\to A')$ on a channel $\mN\in\cptp(A)$ can be represented as
\begin{equation}\label{5}
\Theta^{A\to A'}\left[\mN^A\right]=\e_{\text{post}}^{EA_1\to A_1'}\circ\mN^{A_0\to A_1}\circ\e_{\text{pre}}^{A_0'\to EA_0},
\end{equation}
where $\e_{\text{pre}}\in\cptp(A_0'\to EA_0)$ and $\e_{\text{post}}\in\cptp(EA_1\to A_1')$ are fixed quantum channels corresponding to pre-processing and post-processing of the channel $\mN^A$. This result indicates that superchannels are not just mathematical abstractions, but in fact  have a physical realization.

\begin{figure}[h]
\centering
\includegraphics[width=.4\textwidth]{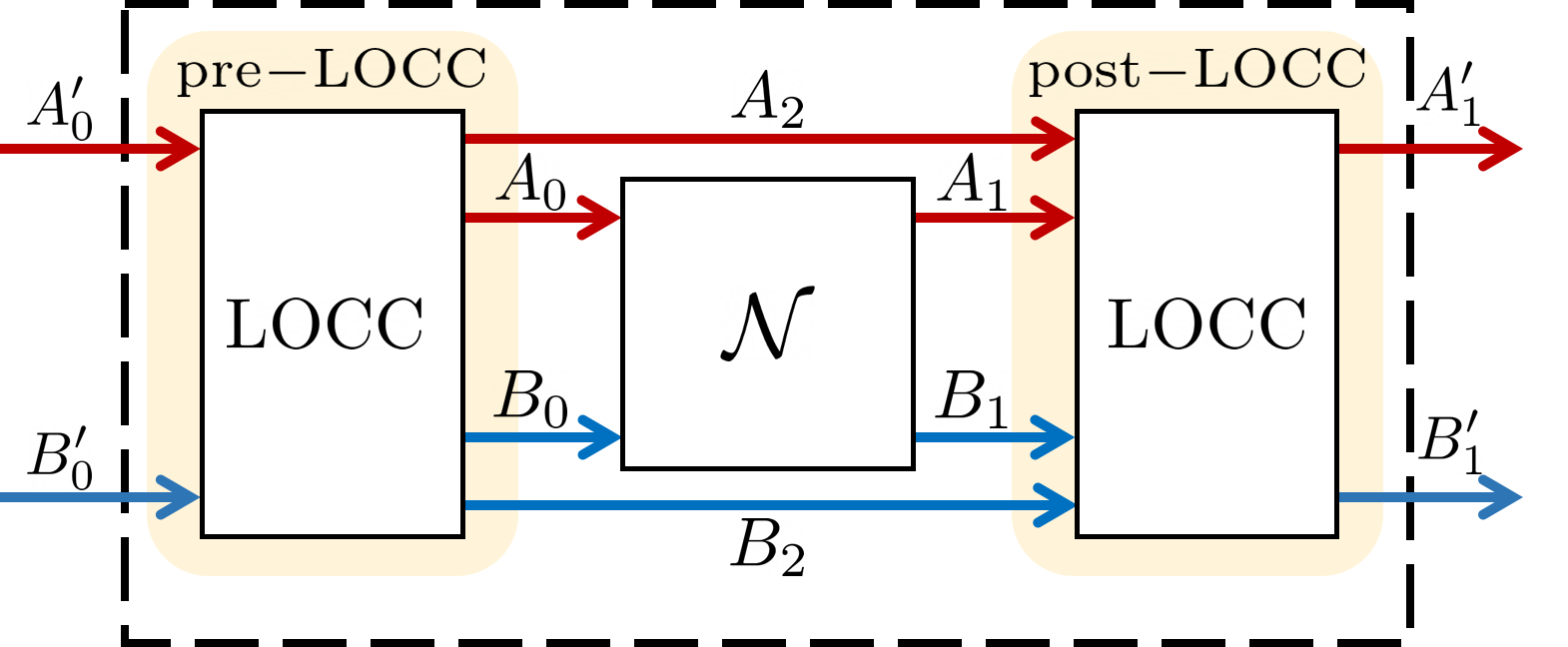}
 \caption{\linespread{1}\selectfont{\small The action of an LOCC superchannel on a bipartite channel $\mN$. }}
 \label{locc}
\end{figure}

Crucial to this work is the set of superchannels that are built using local quantum operations and classical communication (LOCC) \cite{Chitambar-2014a}.  A general LOCC superchannel transforms one quantum channel $\mN^{AB}$ into another by an application of pre- and post- LOCC channels, as depicted in Fig. \ref{locc}.  For given input-output systems, we denote the set of LOCC superchannels as $\locc(AB \to A'B')$.  Throughout this paper, double lines will represent classical systems, whereas single lines are reserved for quantum systems.  Classical channels are denoted by $\mN^{XY}\in\cptp(X_0Y_0\to X_1Y_1)$, and we write $\mN_{\text{CHSH}}^{XY}$ to denote a bipartite classical channel that gives the maximal violation of the CHSH Inequality.  Note that every  classical channel belongs to LOCC.

\section{Quantum Processes}  

\begin{figure}[h]
\centering
\includegraphics[width=.4\textwidth]{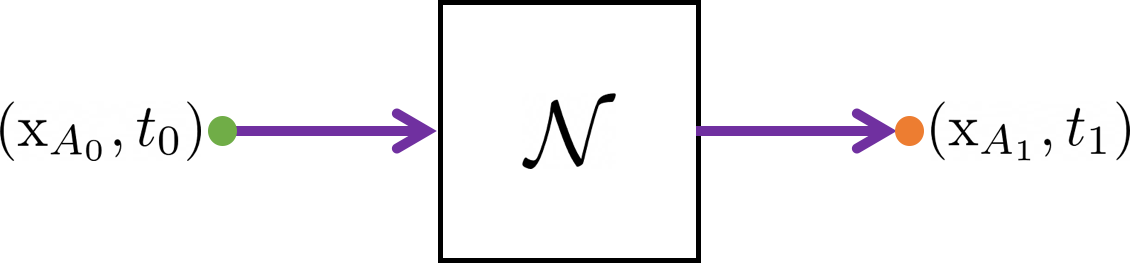}
 \caption{\linespread{1}\selectfont{\small A quantum process $(\mN,\Delta \mbf{x},\Delta t)$ takes an input $\rho$ at arbitrary spacetime point $(\mbf{x}_{A_0},t_0)$ and generates an output $\mN(\rho)$ at spacetime point $(\mbf{x}_{A_1}=\mbf{x}_{A_0}+\Delta\mbf{x},t_1=t_0+\Delta t)$.  The input-output delay time of $\mN$ in this process is $\Delta t$.}}
 \label{Fig:process}
\end{figure}

Mathematically, a quantum channel $\mN\in\cptp(A)$ is a map that sends positive operators of system $A_0$ to positive operators of system $A_1$.  Usually this abstraction is sufficient to probe the foundations and theoretical limitations of quantum information processing.  However, to properly characterize Bell nonlocality in quantum mechanics, we need to understand quantum channels as objects that generate transformations across fixed intervals in space and time, something we broadly refer to as \textit{quantum processes}.  A quantum process, denoted by $(\mN,\Delta\mbf{x},\Delta t)$, is a quantum channel $\mN\in\cptp(A)$ that transforms any state $\rho\in\md(A_0)$ at arbitrary spacetime point $(\mbf{x}_{A_0},t_0)$ into state $\mN(\rho)\in\md(A_1)$ at spacetime point $(\mbf{x}_{A_1},t_1)$, where $\mbf{x}_{A_1}=\mbf{x}_{A_0}+\Delta\mbf{x}$ and $t_1=t_0+\Delta t$ (see Fig. \ref{Fig:process}).  Here we assume for simplicity that systems $A_0$ and $A_1$ are at rest in the same inertial frame and the coordinates are measured with respect to this frame.  The time interval $\Delta t\geq 0$ in a quantum process $(\mN,\Delta\mbf{x},\Delta t)$ is called the \textit{input-output delay time} of $\mN$ (or simply the ``delay time'' of $\mN$), and it quantifies how quickly the channel input propagates to the output.

Some quantum processes can be physically implemented and some cannot. For example, the no-signaling principle of special relativity prohibits $(\mN,\Delta\mbf{x},\Delta t)$ from being physically realizable whenever $\mN$ is able to transmit information and $\Delta t<\Delta\mbf{x}/c$, where $c$ is the speed of light.  Quantum processes with zero input-output delay time are known as \textit{instantaneous}, and they play an important role in this theory.  Whenever $\Delta\mbf{x}>0$, an instantaneous process $(\mN,\Delta\mbf{x},0)$ is physically realizable if and only if $\mN$ is a replacement channel, i.e. $\mN$ has the form $\mN_\rho(X):=\tr[X]\rho$ for $\rho\in\md(A_1)$.  An instantaneous implementation of $\mN_\rho$ is done as follows: knowing that Alice will receive the channel input at time $t_0$, Bob simply prepares the state $\rho$ at $t_0$.  

\begin{figure}[h]
\centering
\includegraphics[width=.45\textwidth]{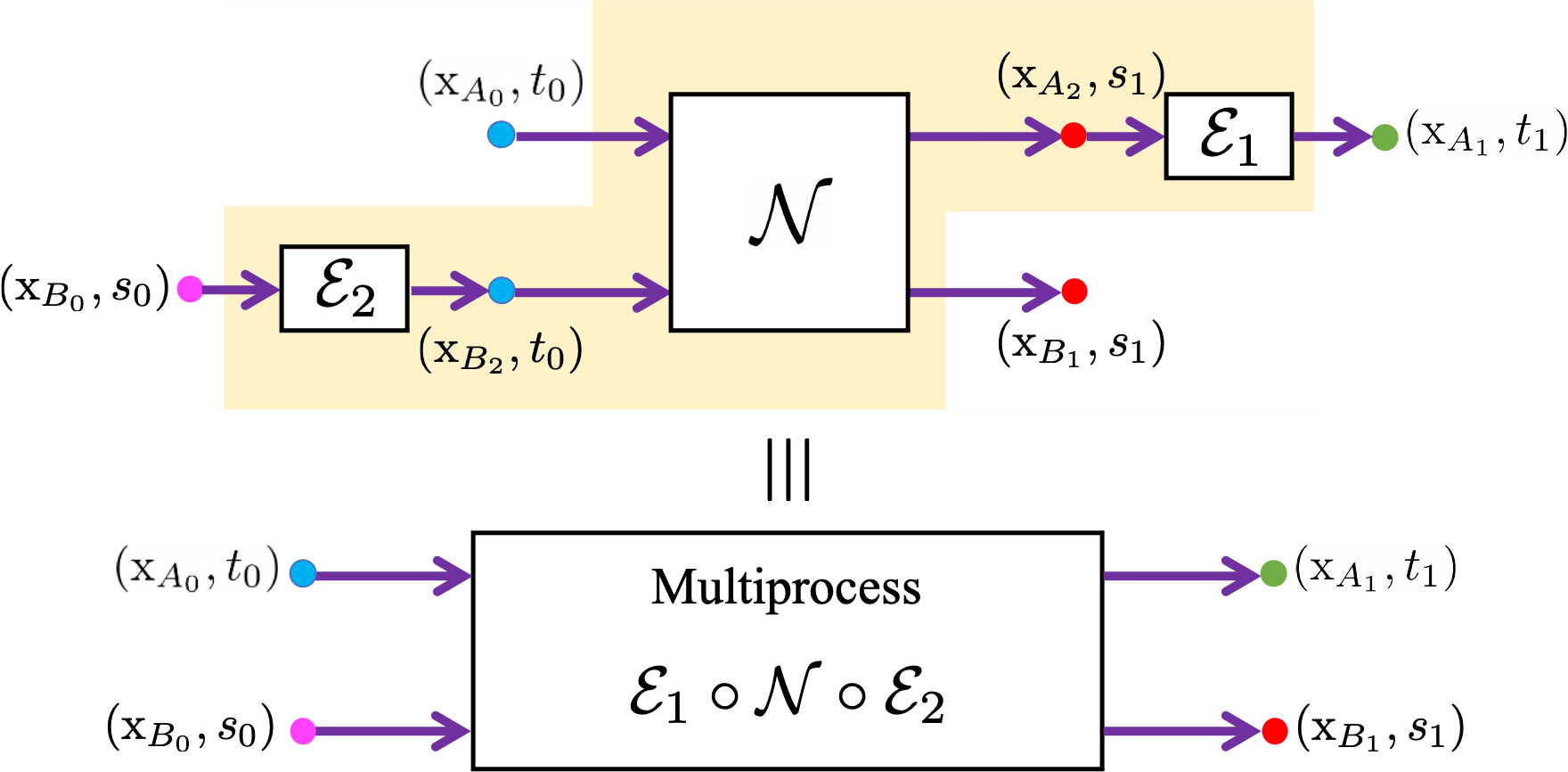}
 \caption{\linespread{1}\selectfont{\small The bipartite channel $\mathcal{E}_2\circ\mathcal{N}\circ\mathcal{E}_1$ can have differing delay times for the different subsystems if it is built by composing multiple processes (top).  The result is a multiprocess (bottom). Dots with the same colour correspond to the same time.
 }}
 \label{Fig:multiprocess}
\end{figure}

The abstraction of quantum processes also applies to bipartite channels $\mN\in\cptp(AB)$.  In this case, however, the input and output of the channel are each distributed across two points in space.   Hence, a bipartite process $(\mN,\Delta \mbf{x}_A,\Delta\mbf{x}_B,\Delta t)$ is the channel $\mN$ that transforms a bipartite state $\rho^{A_0B_0}$ held at $(\mbf{x}_{A_0},t_0)$ and $(\mbf{x}_{B_0},t_0)$ into the state $\mN(\rho^{A_0B_0})$, held at $(\mbf{x}_{A_1},t_1)$ and $(\mbf{x}_{B_1},t_1)$, where $\mbf{x}_{A_1}=\mbf{x}_{A_0}+\Delta\mbf{x}_A$, $\mbf{x}_{B_1}=\mbf{x}_{B_0}+\Delta\mbf{x}_B$, and $t_1=t_0+\Delta t$.  Note that the number of inputs and outputs can differ by treating one of the systems as trivial.  Multipartite processes are defined in the same way: all inputs at spatial coordinates $(\mbf{x}_1,\mbf{x}_2,\cdots,\mbf{x}_n)$ evolve to outputs at spatial coordinates $(\mbf{x}_1+\Delta \mbf{x}_1,\mbf{x}_2+\Delta \mbf{x}_2,\cdots,\mbf{x}_n+\Delta \mbf{x}_n)$ in the \textit{same} time interval $\Delta t$.  By composing processes together, we obtain a multiprocess.  The only difference between a process and a multiprocess is that the latter can have differing time delays for different subsystems due to the composition, whereas the former cannot.  The general idea is depicted in Fig. \ref{Fig:multiprocess}.

\begin{figure}[h]
\centering
\includegraphics[width=.45\textwidth]{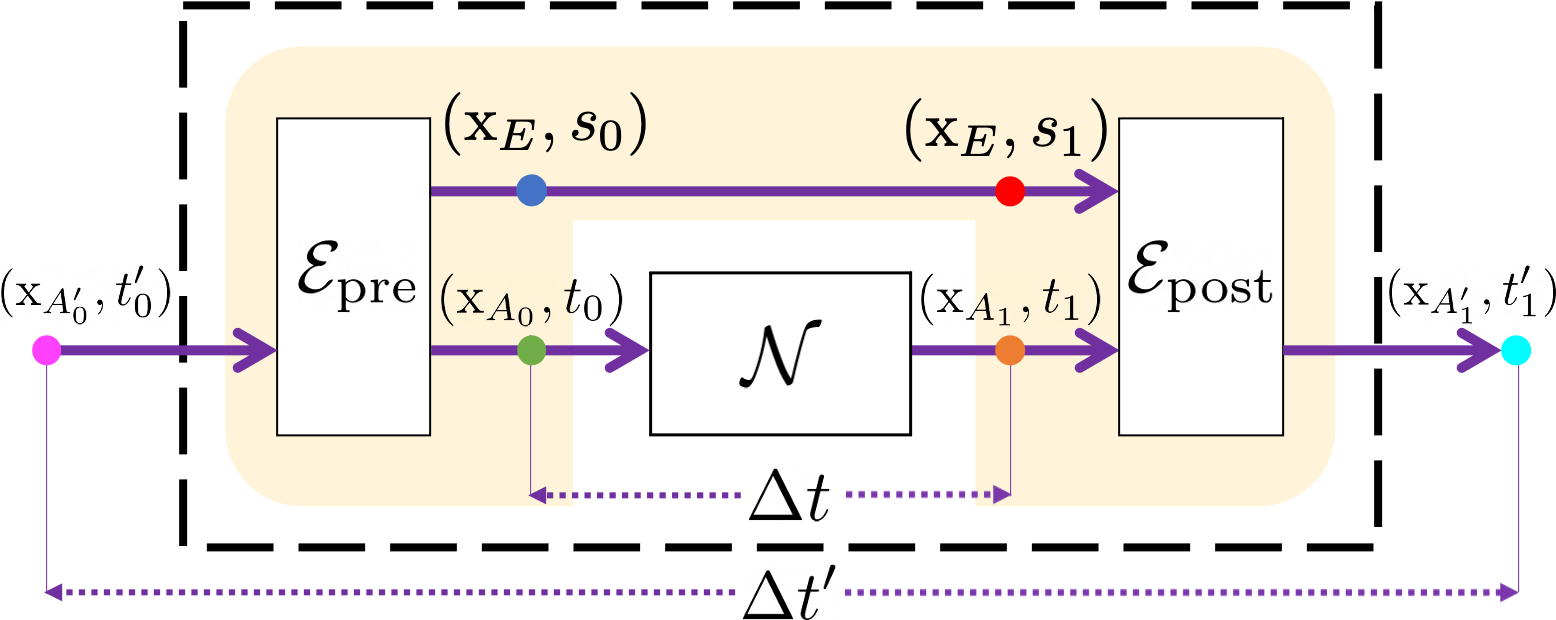}
 \caption{\linespread{1}\selectfont{\small A quantum superprocess $(\Theta,\Delta \mbf{x}\to\Delta \mbf{x}',\Delta t\to \Delta t')$ converts process $(\mN,\Delta \mbf{x},\Delta t)$ to process $(\mN',\Delta \mbf{x}',\Delta t')$.  The shaded yellow area represents the action of the superprocess. Since $\mE_\pree$ and $\mE_\post$ might belong to multiprocess, the times $s_0$ and $s_1$ are not necessarily equal to $t_0$ or $t_1$. For example, in Fig.~\ref{Fig:superprocess-pre} we will see an example in which $t_0<t_0'=s_0=s_1=t_1$.
 }}
 \label{Fig:superprocess}
\end{figure}

We next go one step forward and define a quantum \textit{superprocess} as a superchannel that transforms one process $(\mN,\Delta\mbf{x},\Delta t)$ into another $(\mN',\Delta\mbf{x}',\Delta t')$.  We denote such objects by $(\Theta,\Delta \mbf{x}\to\Delta \mbf{x}',\Delta t\to \Delta t')$.  A superprocess is constructed by invoking Eq.~\eqref{5} and considering $\e_{\text{pre}}$ and $\e_{\text{post}}$ as processes, as they would be in any physical implementation of a superchannel.  A general superprocess then has the form of Fig.~\ref{Fig:superprocess}.  The wires entering and leaving each channel might be multiple systems bundled together. 

If the overall process transformation is $(\mN,\Delta\mbf{x},\Delta t)\to (\mN',\Delta\mbf{x}',\Delta t')$, then there are four possibilities for how $\e_{\text{pre}}^{A_0'\to EA_0}$ and $\e_{\text{post}}^{EA_1\to A_1'}$ can facilitate this transformation.  
\begin{enumerate}
\item The input to $\mN$ at the point $(\x_A,t_0)$ depends on the input to $\e_{\text{pre}}$ at the point $(\x_{A_0'},t_0')$ (and hence $t_0'\leq t_0$)
and the output of $\e_{\text{post}}$ at $(\x_{A_1'}, t_1')$ depends on the output of $\mN$ at $(\x_{A_1},t_1)$ (and hence $t_1\leq t_1'$). In this case the input-output delay time is always non-decreasing, i.e. $\Delta t\leq \Delta t'$.

\item The input to $\mN$ at the point $(\x_A,t_0)$ does not depend on the input to $\e_{\text{pre}}$ at the point $(\x_{A_0'},t_0')$ 
but the output of $\e_{\text{post}}$ at $(\x_{A_1'}, t_1')$ depends on the output of $\mN$ at $(\x_{A_1},t_1)$ (and hence $t_1\leq t_1'$). This means that $\mE_\pree^{A_0'\to A_0E}$ as appear in Fig.~\ref{Fig:superprocess} is a quantum channel whose output at $A_0$ is fixed and independent on the input of the channel. This is a special case of a semi-causal channel~\cite{BGNP2001} and was proven in~\cite{ESW2002} to have for all $\omega\in\md(A_0')$ the form
\be\label{preee}
\e_{\text{pre}}^{A_0'\to EA_0}\left(\omega^{A_0'}\right)=\mL^{A_0'A_2\to E}\left(\rho^{A_0A_2}\otimes\omega^{A_0'}\right)\;,
\ee
where $A_2$ is some auxiliary system, $\rho\in\md(A_0A_2)$ is a fixed quantum state, and $\mL\in\cptp(A_0'A_2\to E)$ is a quantum channel. 
However, the form above implies that $\mL^{A_0'A_2\to E}$ can be ``absorbed" into $\mE_{\post}$ (of Fig.~\ref{Fig:superprocess}) so that w.l.o.g. in Fig.~\ref{Fig:superprocess-pre} we replaced $\mE_\post^{EA_1\to A_1'}\circ\mL^{A_0'A_2\to E}$ with $\mE_\post^{A_0'A_1A_2\to A_1'}$.
Hence when $t_0$ is sufficiently earlier than $t_0'=t_1$, one can attain 
$\Delta t=t_1-t_0> \Delta t'=t_1'-t_0'$.  Note that the difference $\Delta t'$ is determined entirely by the delay time of $\e_{\text{post}}$, whereas $\Delta t$ is the delay time of the input process $(\mN,\Delta\mbf{x},\Delta t)$. 

\begin{figure}[h]
\centering
\includegraphics[width=.48\textwidth]{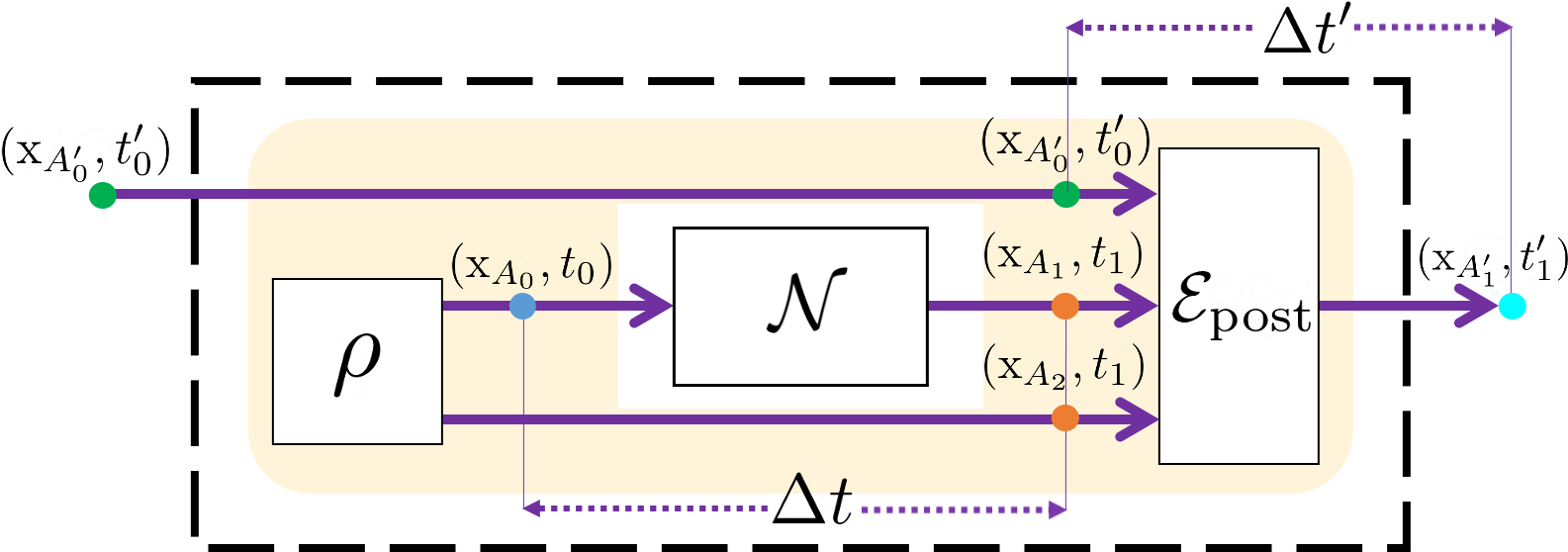}
 \caption{\linespread{1}\selectfont{\small When the input to $\mN$ does not depend on the input to $\mE_{\text{pre}}$  it is possible to construct a superprocess $(\Theta,\Delta \mbf{x}\to\Delta \mbf{x}',\Delta t\to \Delta t')$ that  decrease the input-output delay time, $\Delta t'<  \Delta t$. Observe in particular  that we absorbed $\mL^{A_0'\to E}$ that appears in the expression~\eqref{preee} of $\mE_{\text{pre}}$ into $\mE_{\text{post}}$. In this case the delay time $\Delta t'$ depends only on the implementation of $\mE_{\text{post}}$ and in general can be much smaller than $\Delta t$. Dots with the same colour correspond to the same time.}}
 \label{Fig:superprocess-pre}
\end{figure}

\item The input to $\mN$ at the point $(\x_A,t_0)$ depends on the input to $\e_{\text{pre}}$ at the point $(\x_{A_0'},t_0')$ 
but the output of $\e_{\text{post}}$ at $(\x_{A_1'}, t_1')$ does not depend on the output of $\mN$ at $(\x_{A_1},t_1)$. The superprocess in this case acts as \emph{replacement} superchannel, always outputting the same channel 
\be\label{trivial}
\Theta[\mN]=\e_{\text{post}}^{E\to A_1'}\circ\e_{\text{pre}}^{A_0'\to E}
\ee
 irrespective of the input channel $\mN$. Here,  $\e_{\text{pre}}^{A_0'\to E}\eqdef\tr_{A_0}\circ\e_{\text{pre}}^{A_0'\to EA_0}$ and for all $\rho\in\md(E)$
 $\e_{\text{post}}^{E\to A_1'}(\rho^E)\eqdef \e_{\text{post}}^{EA_1\to A_1'}(\rho^E\otimes\omega^{A_1})$ for some arbitrary fixed $\omega\in\md(A_1)$ . Therefore, while this case might result with $\Delta t>\Delta t'$~\footnote{e.g. remove $\mE_\post$ from Fig.~\ref{Fig:superprocess}, discard the output to $\mN$ at $(\x_{A_1},t_1)$,  and extend $(\x_E,t_0)$ with a straight line to the output $(\x_{A_1'},t_1')$ so that $t_0=t_1'$} it is somewhat a trivial case and will not play an important role in our formalism. 

\item The input to $\mN$ at the point $(\x_A,t_0)$ does not depend on the input to $\e_{\text{pre}}$ at the point $(\x_{A_0'},t_0')$ 
and the output of $\e_{\text{post}}$ at $(\x_{A_1'}, t_1')$ does not depend on the output of $\mN$ at $(\x_{A_1},t_1)$. As in the previous case, also here the superprocess acts as the \emph{replacement} superchannel that always output the  channel given in~\eqref{trivial}. 
\end{enumerate}

\begin{figure}[h]
\centering
\includegraphics[width=.48\textwidth]{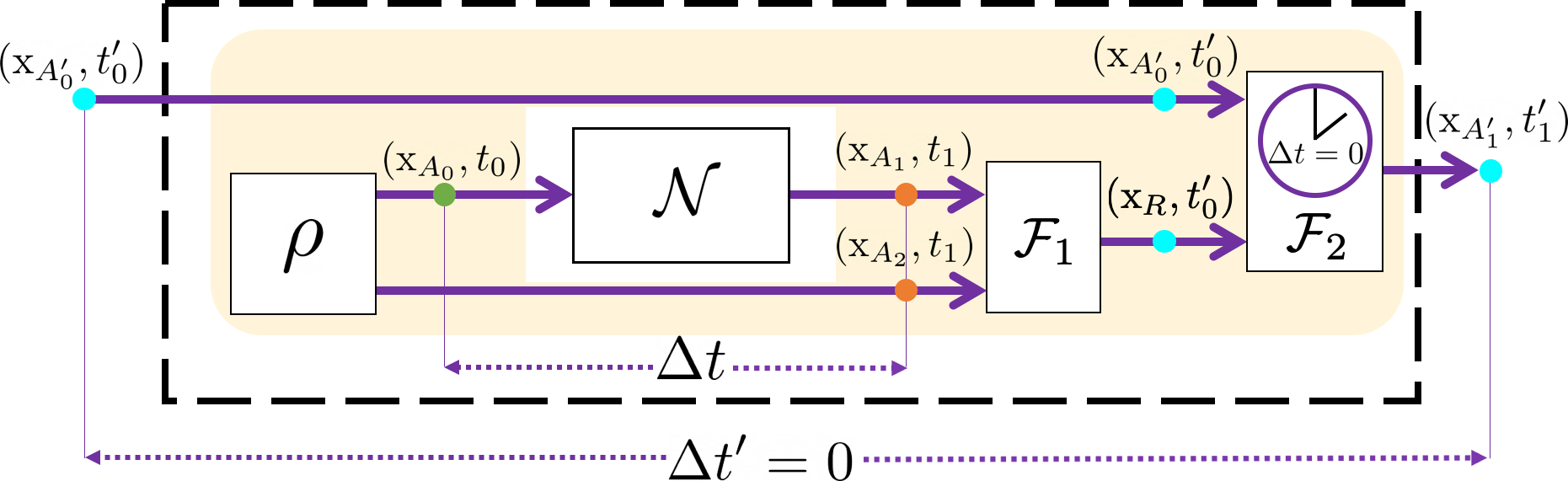}
 \caption{\linespread{1}\selectfont{\small The most general superprocess that can convert a quantum process $(\mN^A,\Delta\mbf{x},\Delta t)$ with $\Delta t>0$ into an instantaneous quantum process. Irrespective of the time delays of $\mN$ and $\mF_1$, since $\mF_2$ is instantaneous we have $t_1'=t_0'\geq t_1> t_0$. In particular, $\Delta t'=t_1'-t_0'=0$.
 }}
 \label{zerodelaytime}
\end{figure}

The superprocess depicted in Fig.~\ref{Fig:superprocess-pre} can also produce $\Delta t'=0$ irrespective of the delay time $\Delta t$ of the quantum process $(\mN,\Delta\mbf{x},\Delta t)$. This happens if $\mE_{\post}$ corresponds to a quantum process with a zero delay time, or more generally, if $\mE_{\post}$ has the form (see Fig.~\ref{zerodelaytime})
\be
\mE_{\post}^{A_0'A_1A_2\to A_1'}=\mF_2^{A_0'R\to A_1'}\circ\mF_1^{A_1A_2\to R}
\ee
where $\mF_1^{A_1A_2\to R}$ corresponds to a quantum process with arbitrary delay time, whereas $\mF_2^{A_0'R\to A_1'}$ corresponds to an instantaneous quantum process. Since $\mF_2$ is instantaneous the delay time $\Delta t'=0$ irrespective of the delay times associated with $\mN$ and $\mF_1$. In Fig.~\ref{zerodelaytime} we depicted the most general superprocess with the property that it converts a non-instantaneous quantum process into an instantaneous one.

To isolate the essential features of this theory, going forward we will assume that the spatial intervals $\Delta \mbf{x}$ and $\Delta\mbf{x}'$ are given and fixed. Then, the only processes we consider are specified by the channel and its delay time, $(\mN,\Delta t)$. The relevant superprocesses in this case are characterized by a superchannel and its induced change in delay times, $(\Theta,\Delta t\to\Delta t')$. 

\section{Entanglement Theory with Instantaneous Resources, alias Bell Nonlocality} 
 
We are interested here in quantum superprocesses that can be realized by performing LOCC.  These are superprocesses $(\Theta,\Delta t\to\Delta t')$ with $\Theta\in \locc(AB\to A'B')$, and they constitute the free operations in this resource theory.  Every LOCC superchannel has the form given in Fig. \ref{locc}, and to isolate the essential features of this approach, we will make the simplifying assumption that all local operations are instantaneous, a reasonable assumption to make when Alice and Bob's laboratories are separated relatively far apart.  Consequently, the delay time $\Delta t$ of an LOCC channel will always be proportional to the number of communication exchanges conducted in the particular implementation of the channel.

The fundamental question is whether one quantum process can be converted to another using an LOCC superprocess.  When $(\mN,\Delta t)\to (\mN',\Delta t')$ is achievable by LOCC, we write
\begin{align}
(\mN,\Delta t)\xrightarrow{\text{\upshape LOCC}}(\mN',\Delta t')\;.
\end{align}
The free objects in this resource theory are the quantum processes that can be generated by LOCC ``from scratch'' (i.e. from the trivial process),
\begin{equation}
(\id^1,0)\xrightarrow{\text{LOCC}}(\mN,\Delta t)\;,
\end{equation}
where $\id^1$ represents here the trivial state/channel; i.e. the only element of $\md(\mbb{C})\cong\cptp(\mbb{C}\to\mbb{C})$.
Notice that every LOCC channel $\mL$ belongs to some free process $(\mL,\Delta t)$ with $\Delta t\in[0,+\infty]$.  As a special case, a bipartite LOCC channel $\mL$ is LOSR if it belongs to an instantaneous LOCC process $(\mL,0)$.  In other words, LOSR corresponds precisely to the family of instantaneous LOCC processes.  

There are non-LOCC bipartite channels $\mathcal{N}$ whose instantaneous process $(\mathcal{N},0)$ can also be physically implemented.  These are channels that belong to the class of local operations and shared entanglement (LOSE).  We say that $\mN\in\lose(AB)$ if there exists some entangled state $\omega\in\md(A_2B_2)$ such that 
 \be\label{lose}
\mN^{AB}=\Upsilon\left[\omega^{A_2B_2}\right]\;,
 \ee
where $\Upsilon$ is an LOSR superchannel comprising of the two local channels $\mE\in\cptp(A_0A_2\to A_1)$ and $\mF\in\cptp(B_0B_2\to B_1)$ such that for any $\rho\in\md(A_0B_0)$ 
  \ba
&\Upsilon\left[\omega^{A_2B_2}\right]\left(\rho^{A_0B_0}\right)\\
&\eqdef\mE^{A_0A_2\to A_1}\otimes\mF^{B_0B_2\to B_1}\left(\rho^{A_0B_0}\otimes\omega^{A_2B_2}\right)\;.
 \ea
 
 \begin{figure}[h]
\centering
\includegraphics[width=.30\textwidth]{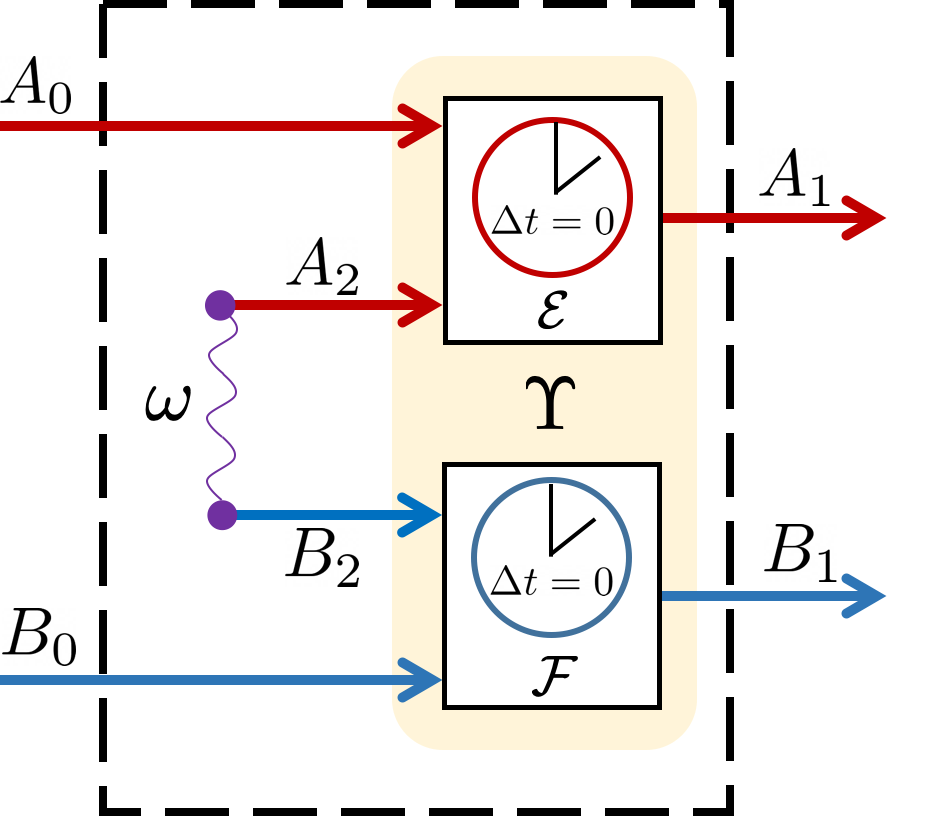}
 \caption{\linespread{1}\selectfont{\small Implementation of an LOSE quantum process with an entangled state $\omega\in\md(A_2B_2)$ and local superprocess $\Upsilon$ that takes the quantum state $\omega^{A_2B_2}$ as its input and then outputs the bipartite quantum channel $\mN^{AB}\eqdef\Upsilon[\omega^{A_2B_2}]$.}}
 \label{fig:lose}
\end{figure}

This framework captures all of entanglement theory.  For instance, all bipartite quantum states (or more generally all bipartite replacement channels) can be implemented with instantaneous processes: the process $(\rho,0)$ with $\rho\in\md(A_1B_1)$ can be viewed as a bipartite quantum process with a one-dimensional input and zero input-output delay time since if Alice and Bob hold $\rho^{A_1B_1}$ and they know they will receive an input at time $t_0$, they can immediately output $\rho^{A_1B_1}$ at time $t_0$.  Hence, entangled states can be implemented with instantaneous processes that are non-LOCC, and separable states can be implemented with the instantaneous LOCC processes (i.e. LOSR).  Converting one entangled state $\rho^{A_1B_1}$ to another $\sigma^{A_1B_1}$ by LOCC then becomes a special case of converting one instantaneous process to another by an LOCC superprocess,
\begin{equation}
\label{Eq:entanglement-transformation}
(\rho^{A_1B_1},0)\xrightarrow{\text{LOCC}}(\sigma^{A_1B_1},0).
\end{equation}

The central thesis of this paper is that the framework of LOCC superprocesses also encompasses the resource theory of quantum Bell nonlocality.  Every experimental test for Bell nonlocality involves the conversion of an entangled bipartite state $\rho^{A_1B_1}$ into a classical channel $\mN^{XY}$, like in Eq. \eqref{Eq:Born} or Eq. \eqref{Eq:pre-LOCC}.  This is sometimes described as transforming a static resource ($\rho^{A_1B_1}$) into a dynamical classical resource ($\mN^{XY}$).  However, this description does not make explicit the resource-theoretic aspect of nonlocality.  While every classical channel can be implemented by LOCC, not every \textit{instantaneous} classical process admits such an implementation.  Hence, in the resource theory developed here, $(\mN^{XY},\Delta t)$ is a free object for $\Delta t$ sufficiently large (which depends on the distance between Alice and Bob), but $(\mN^{XY},0)$ is a resource if $\mN^{XY}$ is non-LOSR, i.e. not admitting a LHV model.  We can now see how Bell nonlocality emerges as just a special type of entanglement.  Entanglement, whether it be static or dynamic, can be defined as any process $(\mN^{AB},\Delta t)$ that cannot be generated by LOCC,
\begin{align}
\text{Entanglement:} \;\;(\text{id}^1,0)\;&\kern.6em\not\kern -.9em\xrightarrow{\text{LOCC}}(\mN^{AB},\Delta t),
\end{align}
and it is the resource in this theory.  Bell nonlocality is just a particular type of this resource in that it is a classical instantaneous process not reachable by LOCC,
\begin{align}
\text{Bell nonlocality:} \;\;(\text{id}^1,0)\;&\kern.6em\not\kern -.9em\xrightarrow{\text{LOCC}}(\mN^{XY},0).
\end{align}
When referring to quantum Bell nonlocality, we mean some Bell nonlocal process that can be obtained from a quantum process using LOCC,
\begin{align}
(\mN^{AB},\Delta t)&\xrightarrow{\text{LOCC}}(\mN^{XY},0)\label{Eq:BNL-process-1}\\
\text{but}\qquad\;\;(\text{id}^1,0)\;&\kern.6em\not\kern -.9em\xrightarrow{\text{LOCC}}(\mN^{XY},0)\label{Eq:BNL-process-2}.
\end{align}
Eq. \eqref{Eq:BNL-process-1} describes a resource conversion just like Eq. \eqref{Eq:entanglement-transformation} except that it considers more general channels and a possible change in input-output delay time.  For example, the quantum Bell nonlocality possessed by the maximally entangled state $\ket{\phi_+^{A_1B_1}}$ (with $|A_1|=|B_1|=2$) can be expressed as the resource conversion
\begin{equation}
(\phi_+^{A_1B_1},0)\xrightarrow{\text{LOCC}}(\mN^{XY}_{\text{CHSH}},0),
\end{equation}
where $\phi_+^{A_1B_1} := | \phi_+ \rangle\langle \phi_+|^{A_1B_1}$.

Let us now examine Eq. \eqref{Eq:BNL-process-1} closer and consider which LOCC superprocesses can convert a quantum process into an instantaneous classical one.  Since the delay time of the target process is zero, there are two possibilities for the delay time $\Delta t$ in the initial process $(\mN^{AB},\Delta t)$.  In the first case, we have $\Delta t=0$ and we need an LOCC superprocess constructed like either Fig. \ref{Fig:superprocess} or Fig. \ref{zerodelaytime} that is instantaneous-preserving.  The other possibility is that $\Delta t >0$ so that the LOCC superprocess decreases the delay time.  As outlined above, every superprocess that decreases delay time has the form of Fig. \ref{zerodelaytime}.  The following theorem characterizes the structure of LOCC superprocesses for each of these cases.

\begin{figure}[h]
\centering
\includegraphics[width=.47\textwidth]{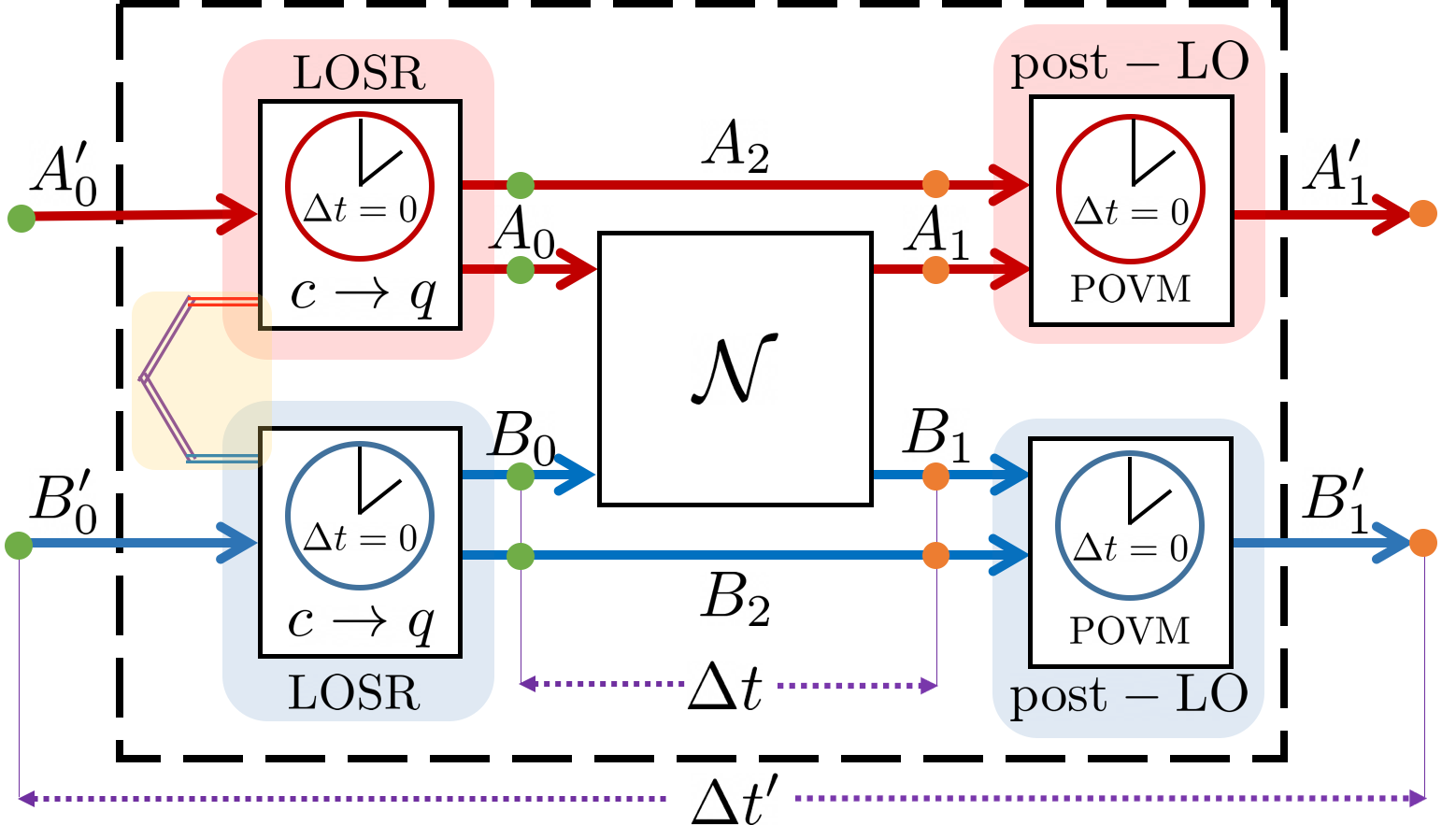}
 \caption{\linespread{1}\selectfont{\small An instantaneous LOCC channel taking the form of LOSR wherein the delay time of the input channel $\mN$ is preserved.}}
 \label{sup_ninst}
\end{figure}

\begin{figure}[h]
\centering
\includegraphics[width=.47\textwidth]{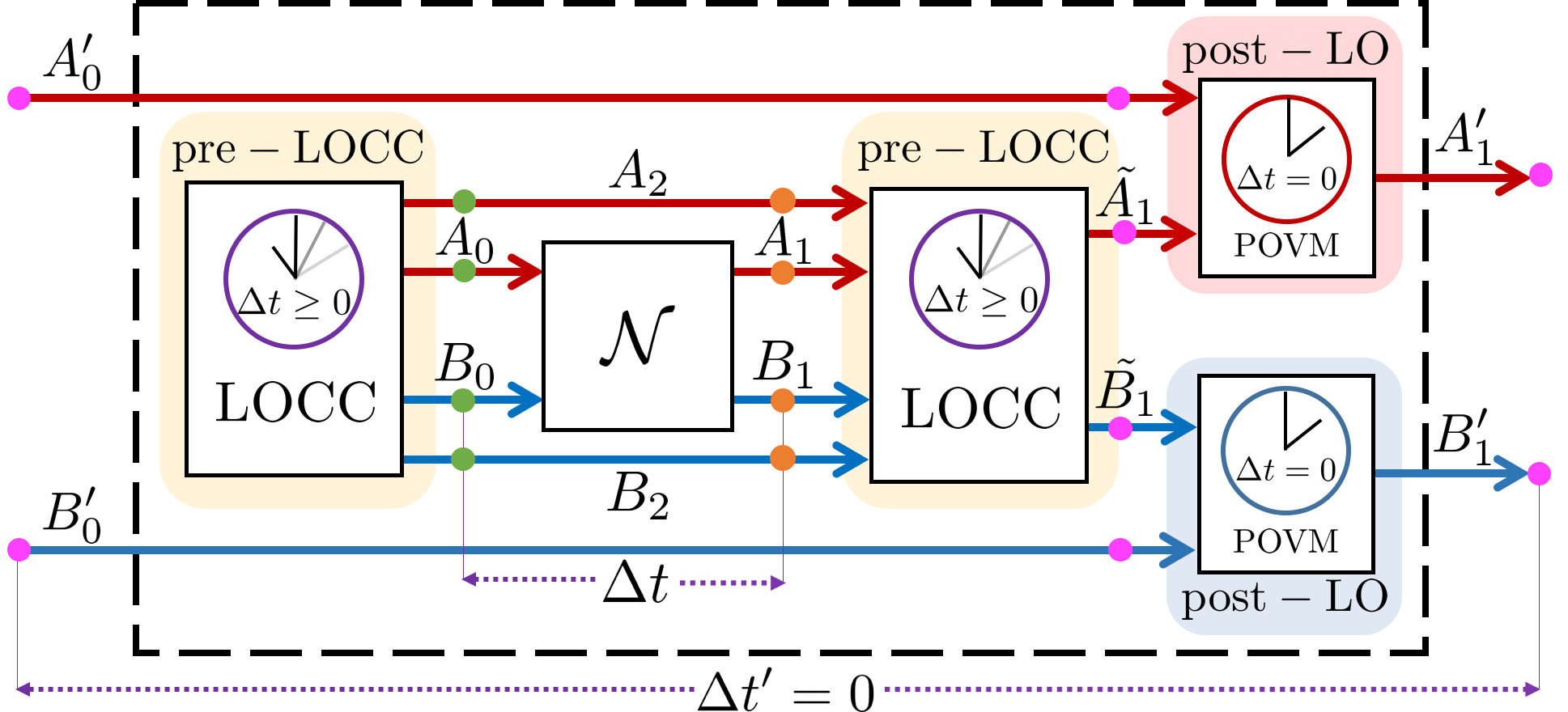}
 \caption{\linespread{1}\selectfont{\small An instantaneous LOCC superchannel containing pre-processing LOCC simulating an instantaneous dynamical resource.}}
 \label{sup_inst}
\end{figure}

\begin{tcolorbox}[enhanced,attach boxed title to top center={yshift=-3mm,yshifttext=-1mm},
  colback=blue!5!white,colframe=blue!75!black,colbacktitle=blue!80!black,
  title=,fonttitle=\bfseries,
  boxed title style={size=small,colframe=blue!50!black} ]
\begin{theorem}
\label{Thm:LOCC-instantaneous}
Let $(\Theta^{AB\to A'B'},\Delta t\to \Delta t')$ be an LOCC superprocess.  
\begin{enumerate}
\item If $\Delta t=\Delta t'=0$, then the superprocess has the form of either Fig.~\ref{sup_ninst} or Fig.~\ref{sup_inst}.
\item If $\Delta t>\Delta t'=0$, then the superprocess has the form of Fig.~\ref{sup_inst}.
\end{enumerate}
\end{theorem}
\end{tcolorbox}

\begin{proof}[Proof of 1]
Recall that \emph{any} superprocess must have the form of  Fig.~\ref{Fig:superprocess}. Since we consider here the bipartite case, we have to replace in Fig.~\ref{Fig:superprocess} the system $A_0$  with $A_0B_0$ and $A_1$ with $A_1B_1$, since the input channel of the superchannel is a bipartite channel $\mN^{AB}$. 
Similarly, since the output of the superchannel $\Theta$ is $A'B'$ we have to replace system $A_0'$ with $A_0'B_0'$ and $A_1'$ with $A_1'B_1'$. Consider now Fig.~\ref{Fig:superprocess} under these replacements.

Recall the four cases that we considered above on how  $\mE_\pree$ and $\e_\post$ can facilitate a superprocess. In the first of these cases we got that $t_0'\leq t_0$ and $t_1\leq t_1'$. Since we assume first that $\Delta t'=\Delta t=0$ we must have that $t_0=t_1=t_0'=t_1'$ so that $\mE_\pree$ and $\mE_\post$ belong to instantaneous quantum processes.  Since we also have that $\Theta$ is LOCC, $\mE_\pree$ and $\mE_\post$ must be both LOCC \emph{and} correspond to instantaneous quantum processes; i.e. both $\mE_\pree$ and $\mE_\post$ are LOSR which is precisely the form given in Fig.~\ref{sup_ninst}. 

In the second case (out of the four cases) the superprocess has the form of Fig.~\ref{Fig:superprocess-pre} adjusted to the bipartite case as discussed above. Now, since $\Delta t'=0$ the superprocess has the form of Fig.~\ref{zerodelaytime} (again adjusted to the bipartite case). Note that since $\Theta$ is LOCC, $\rho$, $\mF_1$, and $\mF_2$ of Fig.~\ref{zerodelaytime} all must be LOCC since we want the superprocess to be both LOCC \emph{and} instantaneous preserving. Hence, $\rho$ is
 is precisely the LOCC process/state that appears on the left side of Fig.~\ref{sup_inst}, $\mF_1$ is the LOCC process that appears at the output of $\mN^{AB}$ in Fig.~\ref{sup_inst}, and since also $\mF_2$ corresponds to instantaneous quantum process it must be an LOSR channel corresponding to the two post-LO channels that appear on the right side on Fig.~\ref{sup_inst}. Note that the shared randomness of $\mF_2$ can be absorbed into the pre-LOCC channel preceding it. Since the third and fourth cases do not lead to new implementations of the superprocess, the proof is concluded for the case $\Delta t'=\Delta t=0$.

Consider now the second part of the theorem in which $\Delta t>\Delta t'=0$. As discussed above, the only implementation (out of the four) of the superprocess for this case is given in Fig.~\ref{zerodelaytime} adjusted to the bipartite case. As discussed in the first part of the proof, an LOCC superprocess of this form is given in Fig.~\ref{sup_inst}. This completes the proof.
 \end{proof}

One of the most important features of this framework is that it provides a physical justification for allowing pre-LOCC processes in experiments of Bell nonlocality.  If the channel $\mN$ is replaced by a quantum state $\rho^{A_1B_1}$, then Fig. \ref{sup_inst} has precisely the same form as Fig. \ref{bell_scenario}.  Thus, hidden Bell nonlocality and superactivation of nonlocality are both features that are operationally accessible in this resource theory.  More precisely, channels having the form of Eq. \eqref{Eq:pre-LOCC} reflect the resource conversion
\begin{equation}
(\rho^{A_1B_1}\otimes\sigma^{A_1'B_1'},0)\xrightarrow{\text{LOCC}}(\mN^{XY},0),
\end{equation}
and so the result of \cite{Masanes-2008a} implies that every entangled state can activate some Bell nonlocal resource using the free operations of this theory.  

Furthermore, all so-called ``anomalies'' of bipartite entanglement \cite{Methot-2007a} vanish under this resource theory since partially entangled states cannot be ``more resourceful'' than maximally entangled ones.  This is because every partially entangled bipartite state can be obtained from a maximally entangled one under LOCC.  Hence whatever process can be freely generated by a weakly entangled state can also be generated by a maximally entangled one.

Let us also remark that Theorem \ref{Thm:LOCC-instantaneous} unifies previous resource-theoretic formulations of Bell nonlocality.  Figure \ref{sup_ninst} describes the transformation of channels by LOSR superchannels.  This model has been studied in the abstract setting of nonlocal ``boxes'' that are governed by a common cause \cite{Wolfe-2020a, Schmid-2020a, Schmid-2020b}.  An alternative operational model has been proposed in Ref. \cite{Gallego-2012a} which is known as wiring and prior-to-input classical communication (WPICC) (see also \cite{Gallego-2017a}).  In the context of quantum resources, WPICC amounts to applying a pre-processing map implementable by LOCC, as in Fig. \ref{sup_inst}.  Our resource theory introduces the notion of input-output delay time as a way to physically motivate both of these types of channel transformations in a unified way.

Returning to Eq. \eqref{Eq:BNL-process-1}, we can consider the range of such transformations.  That is, what instantaneous classical processes can be generated using entanglement as a resource and LOCC superprocessing?  The following theorem shows that dynamical entangled resources (i.e. channels) have no greater ability to generate instantaneous classical processes than static entanglement (i.e. states).

\begin{tcolorbox}[enhanced,attach boxed title to top center={yshift=-3mm,yshifttext=-1mm},
  colback=blue!5!white,colframe=blue!75!black,colbacktitle=blue!80!black,
  title=Channels and States are equi-resourceful,fonttitle=\bfseries,
  boxed title style={size=small,colframe=blue!50!black} ]
\begin{theorem}
Let $(\mN^{AB},\Delta t)$ be a bipartite quantum process with delay time $\Delta t\in[0,\infty]$ and let $(\mM^{XY},0)$ be a bipartite instantaneous classical process.
If $(\mN^{AB},\Delta t)\xrightarrow{\text{LOCC}}(\mM^{XY},0)$, then there exists some bipartite state $\rho^{A_1B_1}$ such that $(\rho^{A_1B_1},0)\xrightarrow{\text{LOCC}}(\mM^{XY},0)$.
\end{theorem}
\end{tcolorbox}

\begin{proof}
We will divide the proof into two parts. Consider first the case that $\Delta t>0$. In this case we saw that the LOCC superprocess must have the form of Fig.~\ref{sup_inst}. Let $\rho^{\tA_1\tB_1}$ be the state at the output $\tA_1\tB_1$ of the second pre-LOCC process in Fig.~\ref{sup_inst}. Then, the proof is concluded with the observation that Fig.~\ref{sup_inst} can be viewed as an LOSR process on the state $\rho^{\tA_1\tB_1}$.

Consider next the case $\Delta t=0$. In this case, the LOCC superprocess can be either the one given in Fig.~\ref{sup_ninst} or the one given in Fig.~\ref{sup_inst}. Since for the latter corresponds to the previous case, we assume now that the supperprocess has the form of Fig.~\ref{sup_ninst}. In this case, since $(\mN^{AB},0)$ is an instantaneous quantum process, it follows that $\mN^{AB}$ is LOSE channel of the form~\eqref{lose}.
By definition, every LOSE channel can be obtained by applying an LOSR superchannel $\Upsilon$ on an entanglement state $\omega^{A_2B_2}$ (see~\eqref{lose}). We therefore conclude that when the LOSR superchannel $\Theta$ of Fig.~\ref{sup_ninst} acts on an LOSE channel $\mN^{AB}=\Upsilon[\omega^{A_2B_2}]$ it produce the channel $\mM^{XY}=\Theta\circ\Upsilon[\omega^{A_2B_2}]$. Since the superchannel $\Theta\circ\Upsilon$ is also LOSR, we  conclude that $\omega^{A_2B_2}$ can be converted to $(\mM^{XY},0)$ with an LOSR superchannel $\Theta\circ\Upsilon$. This concludes the proof.
\end{proof}

\section{Quantification of Bell Nonlocality}  

We next turn to the question of quantifying Bell nonlocality in this resource theory.  A functional ${\rm E} : \cup_{A,B} \cptp(AB) \to \mathbb{R}_+ \cup \{0\}$ was defined in~\cite{gour2019entanglement} as a dynamical entanglement monotone if ${\rm E}(\Theta[\mN])\leq {\rm E}(\mN)$ for any $\mN \in \cptp(AB)$ and for any $\Theta \in \locc(AB \to A'B')$.  Since in our model the resources are quantum processes, we will need to adjust this definition to incorporate the new aspect of input-output delay time associated with each quantum channel.

Monotones of Bell nonlocality for classical channels have been previously explored~\cite{Wolfe-2020a}, where the domain of the functionals presented, is restricted to no-signalling bipartite classical channels. Interestingly, since in our model signalling classical processes are non-instantaneous and therefore free, we do not consider them as well. In fact,  any classical Bell nonlocality monotone will take the zero value on all non-instantaneous classical cprocesses (since classical resources can only be instantaneous). For instance, the Bell nonlocality of the \emph{classical} swap channel, which in previous works were considered as a maximal resource, is zero. For this reason, some monotones of Bell nonlocality that were previously studied in literature need to be adjusted in order to capture the  input-output delay time of the quantum process. 

 Each quantum precess $(\mN^{AB},\Delta t)$ is realizable if the input-output delay time $\Delta t$  is large enough so that the channel $\mN^{A_0B_0\to A_1B_1}$ can be physically implemented (e.g. the swap operation between Alice and Bob cannot be implemented with $\Delta t=0$). For simplicity, we only consider quantum processes with $ \mbf{x}_{A_0}=\mbf{x}_{A_1}\equiv\mbf{x}_A$ and $ \mbf{x}_{B_0}=\mbf{x}_{B_1}\equiv\mbf{x}_B$. On the other hand, the distance between Alice and Bob $\Delta\mbf{x}=\mbf{x}_B-\mbf{x}_A>0$ will be assumed to be strictly positive and kept fixed. Therefore, as in the previous section, there will be no need to include $\Delta\mbf{x}$ in the notation of  quantum processes, and this also enables us, as before, to classify resources as being either instantaneous or non-instantaneous. 
 
 In the following definition we denote by $\mathfrak{Q}(AB)$ the set of all realizable bipartite quantum processes of the form $(\mN^{AB},\Delta t)$  over all $\Delta t\in[0,\infty]$, where $\mN\in\cptp(AB)$. Similarly, we denote by $\mathfrak{Q}(XY)$ the set of all such classical processes $(\mN^{XY},\Delta t)$. Note that any instantaneous resource, $(\mN^{AB},0)\in\mq(AB)$, must be implementable by LOSE.  Otherwise, any communication from Alice to Bob will create a non-zero input-output delay time.

\begin{tcolorbox}[enhanced,attach boxed title to top center={yshift=-3mm,yshifttext=-1mm},
  colback=violet!5!white,colframe=violet!75!black,colbacktitle=violet!65!black,
  title=Dynamical Entanglement Measure,fonttitle=\bfseries,
  boxed title style={size=small,colframe=blue!50!black} ]
\begin{definition}
The function $$E:\bigcup_{A,B}\mq(AB) \to \mbb{R}\;,$$
 where the union is over all finite dynamical systems $A$ and $B$,
 is called a \emph{measure of dynamical entanglement} if on the trivial process $(\id^1,\Delta t)$,
$
E\left(\id^1,\Delta t\right)=0
$,
and for any two bipartite processes $(\mN,\Delta t)$ and $(\mN',\Delta t')$ such that
\be
(\mN,\Delta t)\xrightarrow{\text{\upshape LOCC}}(\mN',\Delta t')
\ee
we have
$
E(\mN,\Delta t)\geq E(\mN',\Delta t')\;.
$
\label{mondef}
\end{definition}
\end{tcolorbox}

When $\mN$ and $\mN'$ are bipartite quantum states the definition above reduces to a measure of (static) entanglement on bipartite states. When $\mN$ and $\mN'$ are classical bipartite channels and $\Delta t=\Delta t'=0$ (i.e. instantaneous processes) the definition above reduces to a measure of Bell non-locality. This means that any measure of dynamical entanglement, $E$, reduces to a measure of Bell non-locality when the domain is restricted to instantaneous classical processes. We denote by $E_{cl}$ the restriction of $E$ to the classical domain, and call it a classical entanglement measure. Note that if $\Delta t>0$ then $E_{cl}(\mN^{XY},\Delta t)=0$ for any classical bipartite channel $\mN\in\cptp(XY)$. Therefore,  we will only consider instantaneous classical processes and write in short $E_{cl}(\mN^{XY})$ for $E_{cl}(\mN^{XY},0)$.

As an example, we introduce a classical Bell nonlocality monotone that is based on channel divergences~\cite{Cooney-2016a,Felix2018,Berta2018,Gour-2019a,Fang2019,Katariya2020,gour2020uniqueness}. Typically, in resource theories such monotones are constructed as the ``distance" (as measured by the divergence) of the resource to the set of free objects. However, in our case, there are two types of free processes: non-instantaneous LOCC quantum processes, and instantaneous LOSR processes. Since divergences are not defined on quantum processes they are insensitive to the distinction between instantaneous vs non-instantaneous resources. For this reason we first define $E$ on classical systems and then extend the domain of $E$ to all channels using the extension techniques developed in~\cite{gour2020optimal}.


Let $D$ be the Kullback-Leibler divergence (relative entropy) defined on any two $n$-dimensional probability vectors $\p$ and $\q$ as
\be
D(\p\|\q)\eqdef\sum_{x=1}^np_x(\log p_x-\log q_x)\;.
\ee
Its extension to classical channels is defined as~\cite{gour2020uniqueness}
$$
 D(\mN\|\mM)\eqdef\max_{x\in[|X_0|]}D\left(\mN\left(|x\lr x|^{X_0}\right)\big\|\mM\left(|x\lr x|^{X_0}\right)\right)\;,
 $$
 for all classical channels $\mN,\mM\in\cptp(X)$. The above function is called a channel divergence since it satisfies the data processing inequality; i.e. for all $\Theta\in\super(X\to Y)$ we have $D\big(\Theta[\mN]\big\|\Theta[\mM]\big)\leq D(\mN\|\mM)$.
 In~\cite{gour2020uniqueness} it was shown that the above classical channel relative entropy is the only channel relative entropy that reduces to the Kullback-Leibler divergence on classical states (i.e. when $|X_0|=1$).
We define the relative entropy of Bell nonlocality as
 \be\label{ecl1}
E_{cl}^{rel}\left(\mathcal{N}^{XY}\right)\eqdef\min_{\mM\in\losr(XY)}D\left(\mN^{XY}\big\|\mM^{XY}\right)\;.
 \ee
 Note that the function above quantifies the dynamical entanglement of an instantaneous bipartite classical channel (i.e. it quantifies quantum Bell non-locality).
 In particular, since the channel divergence $D$ satisfies the DPI it follows that for any $\Theta\in\losr(XY\to X'Y')$
 \be
 E_{cl}^{rel}\left(\Theta\left[\mathcal{N}^{XY}\right]\right)\leq E_{cl}^{rel}\left(\mathcal{N}^{XY}\right)\;,
 \ee
 where $\losr(XY\to X'Y')$ denotes the set of all instantaneous super processes of the form depicted in Fig.~\ref{sup_ninst}.

%


We would like now to extend the above measure to all instantaneous bipartite quantum processes; i.e. to all channels $\mN\in\lose(AB)$ as given in~\eqref{lose}.
We focus in this paper on the quantification of instantaneous resources since the quantification of non-instantaneous entanglement, both static and dynamic, are well understood~\cite{Plenio-2007a,Horodecki-2009a,gour2019entanglement}. Recall that LOSE also include all bipartite quantum states, so that the extensions of $E_{cl}^{rel}$ to all instantaneous bipartite channels, will provide us a way to quantify the quantum Bell nonlocality of a bipartite state. We follow the extension approach recently put forward in~\cite{gour2020optimal} for generalized resource theories.

\begin{tcolorbox}[enhanced,attach boxed title to top center={yshift=-3mm,yshifttext=-1mm},
  colback=violet!5!white,colframe=violet!75!black,colbacktitle=violet!65!black,
  title=Minimal and Maximal Extensions,fonttitle=\bfseries,
  boxed title style={size=small,colframe=blue!50!black} ]
\begin{definition}
Let $$E:\bigcup_{X,Y}\mq(XY) \to \mbb{R}$$
be a classical entanglement measure. Then, for any bipartite quantum channel $\mN \in \lose(AB)$, the minimal and maximal extensions of $E_{cl}$ to all instantaneous quantum processes, denoted by $\overline{E_{cl}}$ and $\underline{E_{cl}}$, respectively, are given by 
\ba\label{ecl2}
&\underline{E_{cl}}(\mathcal{N})\eqdef \sup E_{cl}  \big(\Theta[\mathcal{N}]\big)\quad\text{and}\\
&\overline{E_{cl}}(\mN):= \inf E_{cl}(\calc)
\ea
where the supremum and infimum are taken over all LOCC superchannels $\Theta \in \locc(AB \to XY)$  and all bipartite classical channels $\calc \in \cup_{X,Y}\cptp(XY)$ that satisfy $\mN=\Upsilon[\calc]$ for some superchannel $\Upsilon\in\losr(XY\to AB)$. 
\end{definition}
\end{tcolorbox}

\begin{remark}
Since we are only considering here instantaneous resources, the superchannels $\Theta$ and $\Upsilon$ above are themselves instantaneous LOCC operations. In particular, $\Theta$ has either the form of Fig.~\ref{sup_ninst} (i.e. LOSR) or the form of Fig.~\ref{sup_inst}, while $\Upsilon$ can only have the LOSR form of Fig.~\ref{sup_ninst} since its domain is classical (note that if systems $\tA_1$ and $\tB_1$ of Fig.~\ref{sup_inst} were classical then they would only carry classical shared randomness, and therefore in this case the superchannel depicted in Fig.~\ref{sup_inst} becomes a special case of the LOSR superchannel of Fig.~\ref{sup_ninst}). 
\end{remark}

The following theorem follows directly from the general formalism introduced in~\cite{gour2020optimal} for the extension of resource measures from one domain to a larger one.

\begin{tcolorbox}[enhanced,attach boxed title to top center={yshift=-3mm,yshifttext=-1mm},
  colback=blue!5!white,colframe=blue!75!black,colbacktitle=blue!80!black,
  title=,fonttitle=\bfseries,
  boxed title style={size=small,colframe=blue!50!black} ]
\begin{theorem}
Let $E_{cl}$ be a classical entanglement measure, and $\underline{E_{cl}}$ and $\overline{E_{cl}}$ be as above. Then, $ \underline{E_{cl}}$ and $\overline{E_{cl}}$ are entanglement measures for instantaneous bipartite quantum processes. Moreover, any other such measure $E_{cl}'$ that reduces to $E_{cl}$ on classical instantaneous processes satisfies
$$
\underline{E_{cl}}(\mN) \leq E_{cl}'(\mN)\leq \overline{E_{cl}}(\mN)\quad\forall\;\mN\in\lose(AB)\;.
$$
\label{m_qext}
\end{theorem}
\end{tcolorbox}

Note that from its definition, $\underline{E_{cl}}$ cannot be increased even under pre-LOCC operations as depicted in Fig.~\ref{sup_inst} that results in an instantaneous classical channel. 

\section{Conclusions}  

In this paper we have demonstrated that Bell nonlocality is a property of bipartite quantum systems that can be studied as a special form of entanglement.  This was accomplished by constructing a resource theory based on the abstract notion of a quantum processes.  The important physical element captured in this approach is the input-output time delay of a quantum channel.  Resources in quantum information science can thus be diversely classified as
as (1) classical or quantum, (2) static or dynamic, (3) noisy or noiseless, (4) private or public, and as we add in this paper, (5) instantaneous or non-instantaneous.  


Despite claims that LOSR is the correct operational foundation to study Bell nonlocality in quantum states \cite{Schmid-2020b}, we have established an operationally consistent quantum resource theory in which the inclusion of LOCC pre-processing is physically motivated and follows naturally from the identification of Bell nonlocality as an instantaneous quantum process.  There are also merits to allowing pre-LOCC processing in the resource theory of Bell nonlocality.  First, a well-defined partial order among static and dynamic resource convertibility can be established.  If one allows for LOCC pre-processing, then a maximally entangled state is a resource for generating all forms of quantum correlations, including those present in a Hardy-like Bell test \cite{Hardy-1993a}. More generally, under the use of LOCC pre-processing, maximally entangled states will necessarily have maximal nonlocality with respect to any pre-LOCC monotone.  This reflects our overall conclusion that Bell nonlocality belongs to the same resource theory as quantum entanglement. 

Furthermore, pre-LOCC is able to identify a source of Bell nonlocal states whereas LOSR cannot.  As argued in Ref. \cite{Theurer-2019a}, any physically meaningful quantum resource theory should allow for the possibility of detecting resource states using the free operations of the theory along with the possible consumption of some auxiliary resource state.  LOSR fails to satisfy this criterion while pre-LOCC does not.  Suppose Alice and Bob have access to an i.i.d. source of some state $\rho^{AB}$ which is promised to be a particular Bell nonlocal state.  Using LOSR, they cannot distinguish this state from the product state $\rho^A\otimes\rho^B$.  On the other hand, when using pre-LOCC, measurement outcomes can be discussed between the processing of each copy of $\rho^{AB}$, and by performing a suitable Bell test, they can verify that $\rho^{AB}$ is indeed a resource.
 
We also introduced a systematic way to quantify the dynamical entanglement of quantum processes. As an example, we introduced a measure of instantaneous entanglement (e.g. quantum Bell non-locality) that is based on the   
the Kullback-Leibler divergence. Since the latter is the only asymptotically continuous divergence~\cite{Matsumoto2018} it has numerous applications in quantum information. Remarkably, its extension to classical channels is unique~\cite{gour2020uniqueness}, and therefore, we expect that our specific construction of an instantaneous entanglement measure in~\eqref{ecl1}, and its quantum extensions in~\eqref{ecl2}, will have operational meaning; we leave it for future work.

\bibliographystyle{apsrev4-1}
\bibliography{reference2}

\clearpage

\onecolumngrid
\begin{center}
    \textbf{\large{Appendix}}
\end{center}
\setcounter{equation}{0}

\section*{Additional Results}

\subsection{Fully Bell local}


So far, we have captured the essence of quantum Bell nonlocality and entanglement theory in the framework of LOCC superprocesses. A quantum superprocess, $(\Theta^{AB \to A'B'}, \Delta t \to \Delta t')$, takes a quantum process $(\mathcal{N}^{AB}, \Delta t)$ to another process $(\mathcal{M}^{A'B'}, \Delta t')$. A quantum superprocess  is LOCC when the underlying superchannel $\Theta$ has an LOCC realization, as shown in Fig.~\ref{locc}. While the framework of superprocesses encapsulates all of dynamical entanglement theory, Bell nonlocality emerges out from two restrictions on these superprocesses; i) the resultant channel is classical, i.e. $A'B'=XY$, ii) the superprocess has either of the forms shown in Fig.~\ref{sup_ninst} and~\ref{sup_inst}; 

In this section, we consider delay time preserving (DTP) LOCC (i.e. LOSR)~\footnote{Note that if $\Delta t=0$, then both the superprocesses in Fig.\ref{sup_ninst} and~\ref{sup_inst} preserve the time delay. However, if $\Delta t > 0$ then only the superprocesses of the form as shown in Fig.\ref{sup_ninst} preserve the delay time of the input process, while the superprocesses of the form as shown in Fig.~\ref{sup_inst} anihilates it (see Theorem~\ref{Thm:LOCC-instantaneous}).} superprocesses which take bipartite quantum processes to a bipartite quantum to classical (POVM) processes, i.e. superprocesses of the form $(\Theta^{AB \to A'B'}, \Delta t \to \Delta t)$, where $\Theta$ has the form shown in Fig.~\ref{sup_ninst} with $A'_1$ and $B'_1$ taken to be classical systems. We show that every entangled bipartite channel can be used to simulate, via DTP LOCC superprocesses, at least one bipartite nonlocal POVM channel. Therefore, even on not requiring the output channel to be classical, the nonlocal features of entanglement can still be observed. We call separable (i.e. LOSR) processes \emph{fully Bell local}.




\begin{tcolorbox}[enhanced,attach boxed title to top center={yshift=-3mm,yshifttext=-1mm},
  colback=violet!5!white,colframe=violet!75!black,colbacktitle=violet!65!black,
  title=Fully Bell Local Processes,fonttitle=\bfseries,
  boxed title style={size=small,colframe=blue!50!black} ]

\begin{definition}
Let $(\mathcal{N}^{AB},\Delta t)$ be a a bipartite quantum process with delay time $\Delta t \in [0, \infty]$. $(\mathcal{N}^{AB},\Delta t)$ is said to be \emph{fully Bell local} if for every DTP LOCC (i.e. LOSR) superprocess $(\Theta^{AB \to A'B'},\Delta t \to \Delta t)$, of the form given in Fig.~\ref{sup_ninst}, where $A'_1$ and $B'_1$ are restricted to classical systems, the process $(\Theta[\mathcal{N}], \Delta t)$ is LOSR.
\end{definition}
\end{tcolorbox}
This means that every bipartite POVM process that can be simulated by applying a DTP LOCC superprocess on a fully Bell local process must also be local. We characterize the set of fully Bell local processes in our next theorem.


\begin{figure}[h]
\centering
\includegraphics[width=1\textwidth]{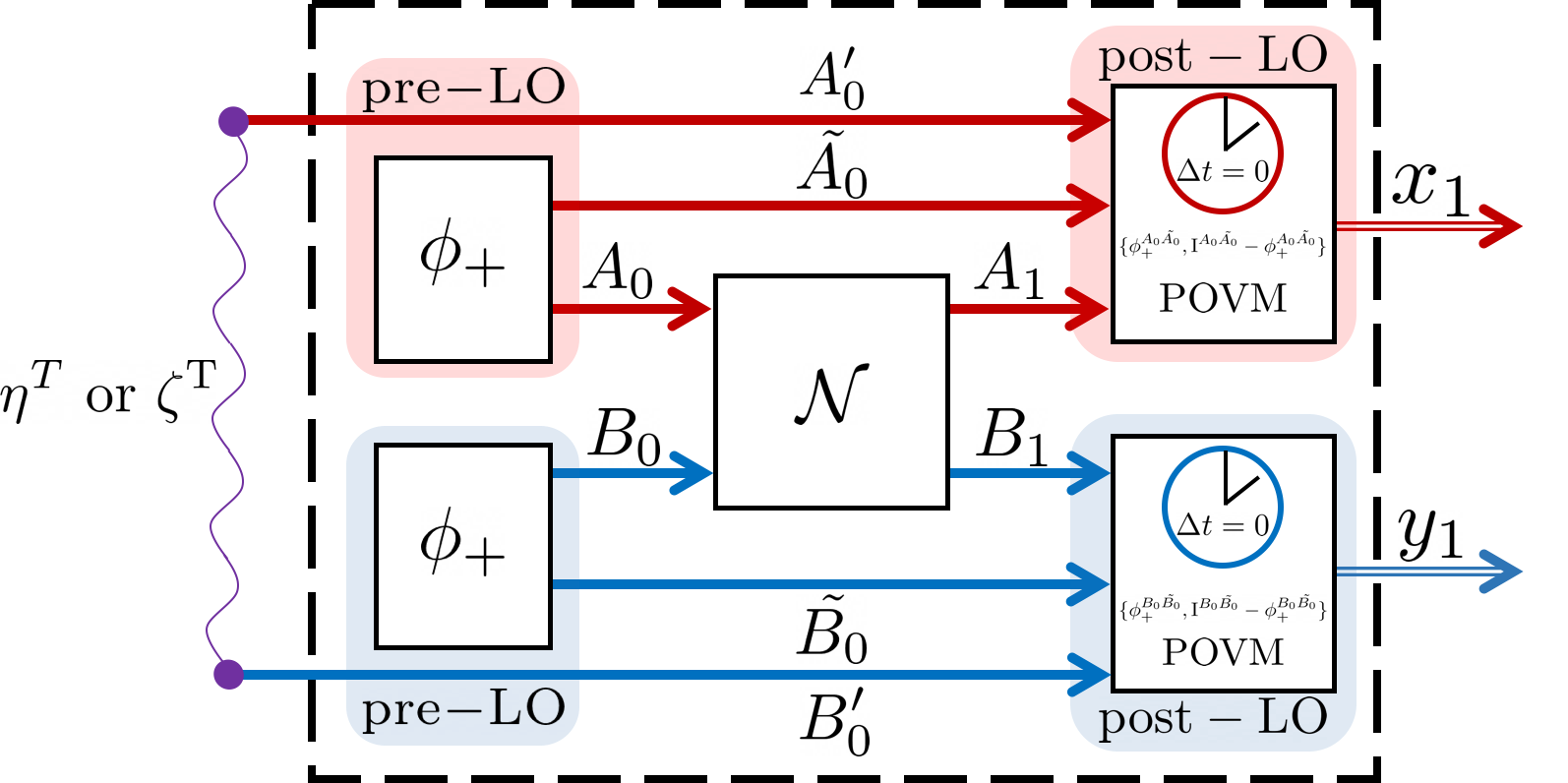}
 \caption{\linespread{1}\selectfont{\small }}
 \label{Thm4_proof}
\end{figure}

\begin{tcolorbox}[enhanced,attach boxed title to top center={yshift=-3mm,yshifttext=-1mm},
  colback=blue!5!white,colframe=blue!75!black,colbacktitle=blue!80!black,
  title=Instantaneous LOCC channels are Fully Bell Local,fonttitle=\bfseries,
  boxed title style={size=small,colframe=blue!50!black} ]
\begin{theorem}
Let $(\mN^{AB}, \Delta t)$ be a bipartite quantum process and $(\mathcal{L}^{AB},0)$ denote the set of all instantaneous LOCC processes from systems $A_0,B_0$ to systems $A_1,B_1$. $(\mN^{AB}, \Delta t)$  is fully Bell local if and only if $(\mN^{AB}, \Delta t) \in (\mathcal{L}^{AB},0)$. Alternatively,  $(\mN^{AB}, \Delta t)$ is fully Bell local if $\mN^{AB}$ admits the form
\begin{equation}
    \mN^{AB} = \sum_j t_j \mE^A_j \otimes \mF^B_j,
\end{equation}
 where $\mE^A_j \in \cptp(A)$ and $\mF^B_j \in \cptp(B)$ for all $j \in \mathbb{R}$ and $t_j$ is a probability distribution. 
\label{fully_Bell_local}
\end{theorem}
\end{tcolorbox}

\begin{proof}

We start by noting that the C-J matrix of a bipartite channel is separable if and only if it is an LOSR channel (i.e. instantaneous LOCC). Clearly, if $\mN$ is LOSR, it is fully Bell local, since an LOSR superchannel takes bipartite LOSR channels to bipartite LOSR channels. We therefore assume that $\mN$ is fully Bell local, and by contradiction, also non-LOSR, i.e. its C-J matrix is not separable. Since every C-J matrix can be viewed as an unnormalized density matrix, for any non-separable C-J matrix $J_\mN$ of a non-LOSR bipartite quantum channel $\mN$, there will always be a witness $W \in \herm(AB\tilde{A}\tilde{B})$, such that $\tr[J_\mN W] < 0$ and $\tr[J_\mM W] \geq 0$ for every LOSR channel $\mM$. Here $\herm (AB\tilde{A}\tilde{B})$ denotes the set of hermitian operators in $\mathfrak{B}(AB\tilde{A}\tilde{B})$. 

Now, express $W=r\eta-t\zeta$, where $\eta,\zeta\in\md(A_0'B_0')$ and $r,t\geq 0$, such that $|A_0'| =|A_0||A_1|$ and $|B_0'| =|B_0||B_1|$. Consider the quantum to classical (qc) channel, $\mF\in\cptp(A_0'B_0'\to X_1Y_1)$, generated by the application of an LOSR superchannel with pre-processing which generates maximally entangled state and a post-processing which includes local projective measurements as shown in Fig.~\ref{Thm4_proof} .
Hence, there will be some outcome in $(x,y)$ in which they both project onto the maximally entangled state. For this outcome, on the inputs $\eta^T$ and $\zeta^T$ the channel satisfies
\begin{equation}
\begin{split}
{\rm p}(x,y|\eta^T)&\eqdef\la x,y|\mF^{AB\to XY}(\eta^T)|x,y\ra\\
&= \frac{1}{|A_0||B_0|}\la\phi_{+}^{A_0'\tilde{A_0'}}\otimes\phi_{+}^{B_0'\tilde{B_0'}}|\mN^{\tilde{A_0}\tilde{B_0}\to A_1B_1} (\phi_{+}^{A_0\tA_0}\otimes\phi_{+}^{B_0\tB_0})\otimes(\eta^{A_0'B_0'})^T|\phi_{+}^{A_0'\tilde{A_0'}}\otimes\phi_{+}^{B_0'\tilde{B_0'}}\ra\\
&=\frac{1}{|A_0||B_0|}\la\phi_{+}^{A_0'\tilde{A_0'}}\otimes\phi_{+}^{B_0'\tilde{B_0'}}|J_\mN \otimes(\eta^{A_0'B_0'})^T|\phi_{+}^{A_0'\tilde{A_0'}}\otimes\phi_{+}^{B_0'\tilde{B_0'}}\ra\\
&=\tr[\rho_{J_\mN}\eta^{A_0'B_0'}],\\
\end{split}
\end{equation}

where all $\phi_+$ are unnormalized and $\rho_{J_\mN} \eqdef \frac{1}{|A_0||B_0|} J$ and similarly ${\rm p}(x,y|\zeta^T)=\tr[\rho_{J_\mN}\zeta^{A_0'B_0'}]$. Therefore,
\be
r{\rm p}(x,y|\eta^T)-t{\rm p}(x,y|\zeta^T)=\tr[\rho_{J_\mN}W]<0\;.
\ee
On the other hand, for any LOSR channel $\mE\in\losr(A_0'B_0'\to X_1Y_1)$ with POVM elements $\{\Pi_{x|i}^{A_0'}\otimes\Pi_{y|i}^{B_0'}\}$ and prior $\{\rm{p}_i\}$ such that $\mE(\cdot)=\sum_i {\rm p}_i\tr\left[(\cdot)\left(\Pi_{x|i}^{A_0'}\otimes\Pi_{y|i}^{B_0'}\right)\right]|xy\lr xy|^{X_1Y_1}$, we have for \emph{any} $x$ and $y$,
\be
r {\rm p}(x,y|\eta^T)-t{\rm p}(x,y|\zeta^T)=\sum_i{\rm p}_i\tr\left[W^T\left(\Pi_{x|i}^{A_0'}\otimes\Pi_{y|i}^{}B_0'\right)\right]=\tr\left[W^T\sum_i{\rm p}_i\Pi_{x|i}^{A_0'}\otimes\Pi_{y|i}^{B_0'}\right]\geq 0,
\ee
where the last inequality follows from the separability of the POVM. Hence, we get a contradiction. This completes the proof. While writing this paper, the authors became aware of an alternative proof provided in~\cite{Schmid2020typeindependent}.
\end{proof}


{\it Remark-}
As a special case of the theorem above, one obtains that a quantum state $\rho\in\md(A_1B_1)$ is fully Bell local (i.e. it only results in Bell local qc channels) if and only if it is separable, that is when $\mN$ is taken to be a replacement channel (i.e. $\mN(\sigma) = \rho$ for all $\sigma \in \md(A_0B_0)$).

\subsection{CHSH Witness}

\begin{tcolorbox}[enhanced,attach boxed title to top center={yshift=-3mm,yshifttext=-1mm},
  colback=blue!5!white,colframe=blue!75!black,colbacktitle=blue!80!black,
  title=Entanglement Witness for Bipartite POVMs,fonttitle=\bfseries,
  boxed title style={size=small,colframe=blue!50!black} ]
\begin{theorem}
Let $\psi_{x_0} \in \md{(A_0)}$ and $\phi_{y_0} \in \md{(B_0)}$ be pure quantum states, $x_0,y_0 \in \{0,1\}$, $x_1,y_1 \in \mathbb{Z}$ and $\oplus$ denote addition modulo 2. Then, the hermitian matrix $W = \sum_{x_1,y_1}W_{x_1y_1} \otimes |x_1y_1\rangle\langle x_1y_1|$ is an $\losr$ witness for $\mathrm{POVM}$ channels in finite dimensions,  where
\be
    W_{x_1y_1} = \sum_{x_0,y_0} \left(\frac{3}{16} - \delta_{x_1 \oplus y_1 = x_0y_0}\right) ( \psi_{x_0}  \otimes \phi_{y_0} ).
\ee
\label{witness}
\end{theorem}
\end{tcolorbox}

\noindent This witness is a generalization of of the CHSH inequality to the case of bipartite POVM channels and is independent of the choice of basis.
\begin{proof}
 Let $\mathcal{E}_{qc}$ be a $\povm$ channel with $\{\Pi_{x_1y_1|x_0y_0}\}_{x_1,y_1}$ as the POVM elements satisfying $\sum_{x_1,y_1} \Pi_{x_1y_1|x_0y_0}= I^{A_0B_0}$.

\begin{equation}
\begin{split}
 \tr\left[W J_{\mathcal{E}_{qc}}\right]& = \sum_{x_1,y_1} \tr\left[ \Pi^{A_0B_0}_{x_1y_1|x_0y_0} \otimes |x_1y_1\rangle\langle x_1y_1|) \ \left(W_{x_1y_1} \otimes |x_1y_1\rangle\langle x_1y_1|\right)\right]   
 \\ &= \sum_{x_1,y_1} \tr\left[ \left(\Pi^{A_0B_0}_{x_1y_1|x_0y_0} \otimes |x_1y_1\rangle\langle x_1y_1|\right) \sum_{x_0,y_0} \left(\frac{3}{16} - \delta_{x_1 \oplus y_1 = x_0y_0}\right) [\psi_{x_0}^A \otimes \phi_{y_0}^B] \otimes |x_1y_1\rangle\langle x_1y_1|\right]
 \\ & = \sum_{x_1,y_1,x_0,y_0} \tr \left[\Pi^{A_0B_0}_{x_1y_1|x_0y_0} (\psi_{x_0}^A \otimes \phi_{y_0}^B) \left(\frac{3}{16} - \delta_{x_1 \oplus y_1 = x_0y_0}\right)\right] \tr\left[|x_1y_1\rangle\langle x_1y_1|\right] \\
 &= \sum_{x_1,y_1,x_0,y_0} {\rm p}(x_1,y_1|x_0,y_0) \left(\frac{3}{16} - \delta_{x_1 \oplus y_1 = x_0y_0}\right) \\
\end{split}
\label{eq20}
\end{equation}
 Now, if $\mathcal{E}_{qc}$ is an $\losr$ channel, then $\Pi_{x_1y_1|x_0y_0} = \sum_{\lambda} t_\lambda \Pi_{x_1|x_0\lambda} \otimes \Pi_{y_1|y_0\lambda}'$ where $\Pi_{x_1|x_0\lambda}$ and $\Pi_{y_1|y_0\lambda}'$ are individual $\povm$ elements and they satisfy $\sum_{x_1} \Pi_{x_1|x_0\lambda}=I^A_0$ and  $\sum_{y_1} \Pi_{y_1|y_0\lambda}'=I^B_0$ and $t_{\lambda}$ is a probability distribution function depending on the random variable $\lambda$. For such a channel, we have 
\begin{equation}
\begin{split}
 \tr\left[W J_{\mathcal{E}_{qc}}\right]& = \sum_{x_1,y_1,\lambda} \tr\left[ t_\lambda (\Pi_{x_1|\lambda} \otimes \Pi_{y_1|\lambda}' \otimes |x_1y_1\rangle\langle x_1y_1|) \ (W_{x_1y_1} \otimes |x_1y_1\rangle\langle x_1y_1|)\right] 
\\ &= \sum_{x_1,y_1,x_0,y_0,\lambda} \ t_\lambda \  {\rm p}_{x_1|\lambda x_0}\  {\rm q}_{y_1|\lambda y_0}\left(\frac{3}{16} - \delta_{x_1 \oplus y_1 = x_0y_0}\right)  \\
 &= \sum_{x_1,y_1,x_0,y_0} {\rm p}_{l}(x_1,y_1|x_0,y_0) \  \left(\frac{3}{16} - \delta_{x_1 \oplus y_1 = x_0y_0}\right) \geq 0,
\label{eq20}
\end{split}
\end{equation}
where $\pr_l$ represents a Bell local probability distribution, ${\rm p}_{x_1|\lambda x_0} = \tr[\Pi_{x_1|\lambda}^A \psi_{x_0}]$ and ${\rm q}_{y_1|\lambda y_0}= \tr[{\Pi'}_{y_1|\lambda}^B \phi_{y_0}]$. This is true for any choice of $\povm$ elements $\Pi_{x_1|x_0\lambda}$ and $\Pi_{y_1|y_0\lambda}'$ and is also independent on the choice of $\psi_{x_0}$ and $\phi_{y_0}$. On the other hand, if $\mathcal{E}_{qc}$ is not an $\losr$ channel, $\Pi_{x_1y_1|x_0y_0}$ cannot admit the convex distribution above. One possible way of constructing such a channel is with the help of a bipartite quantum state which is not fully Bell local (Theorem~\ref{fully_Bell_local}), i.e. entangled. However, from the setup of the CHSH game for entangled states, it is always possible to construct at least one classical channel via the action of a quantum to classical superchannel such that the underlying probability distribution helps the parties to win the game. In other words, for every Bell nonlocal bipartite state, there exists at least one choice of  $\{\Pi_{x_1y_1|x_0y_0}\},\{\psi_{x_0}\}$ and $\{\phi_{y_0}\}$ such that,
\begin{equation}
    \begin{split}
 \tr\left[W J_{\mathcal{E}_{qc}}\right]& = \sum_{x_1,y_1} \tr\left[ \Pi^{A_0B_0}_{x_1y_1|x_0y_0} \otimes |x_1y_1\rangle\langle x_1y_1|) \ (W_{x_1y_1} \otimes |x_1y_1\rangle\langle x_1y_1|)\right] 
 \\ &=\sum_{x_1,y_1,x_0,y_0}{\rm p}_{nl}(x_1,y_1|x_0,y_0)\left(\frac{3}{16} - \delta_{x_1 \oplus y_1 = x_0y_0}\right) < 0, 
    \end{split}
\end{equation}
where $\rm{p}_{nl}$ represents a Bell nonlocal probability distribution.

\end{proof}

\end{document}